\documentclass[aps,pre,showpacs,floats,preprint]{revtex4}
\usepackage{amssymb}
\usepackage{graphics}
\usepackage[dvipdfm]{graphicx}
\usepackage{epsfig}
\usepackage{dcolumn}
\usepackage{amsmath}

\begin{document}

\title{Higher-Order Generalized Hydrodynamics of Carriers and Phonons in
Semiconductors in the Presence of Electric Fields: Macro to Nano}
\author{Cl\'{o}ves G. Rodrigues$^{1}$, A. Rubens B. Castro$^{2,3}$, Roberto Luzzi$^{2}$\footnote{group home
page: www.ifi.unicamp.br/aurea; email: cloves@pucgoias.edu.br}}
\affiliation{$^{1}$Departamento de F\'{\i}sica, Pontif\'{\i}cia
Universidade Cat\'{o}lica de Goi\'{a}s, 74605-010 Goi\^{a}nia, Goi\'{a}s, Brazil\\
$^{2}$Condensed Matter Physics Department, Institute of Physics ``Gleb Wataghin''\\
State University of Campinas-Unicamp, 13083-859 Campinas, SP, Brazil\\
$^{3}$Brazilian Synchrotron Light Laboratory, Campinas, SP, Brazil}

\date{\today }

\begin{abstract}
It is analyzed the hydrodynamics of carriers (charge and heat
motion) and phonons (heat motion) in semiconductors in the presence
of constant electric fields. This is done in terms of a so-called
Higher-Order Generalized Hydrodynamics (HOGH), also referred to as
Mesoscopic Hydro-Thermodynamics (MHT), that is, covering phenomena
involving motions displaying variations short in space and fast in
time and being arbitrarily removed from equilibrium, as it is the
case in modern electronic devices. The particular case of a MHT of
order 1 is described, covering wire samples from macro to nano
sizes. Electric and thermal conductivities are obtained. As the size
decreases towards the nanometric scale, the MHT of order 1 produces
results that in some cases greatly differ from those of the usual
hydro-thermodynamics. The so-called Maxwell times associated to the
different fluxes present in MHT are evidenced and analyzed; they
have a quite relevant role in determining the characteristics of the
motion.
\end{abstract}


\maketitle


\section{Introduction}

The modern advanced technologies, and its resulting end use for
improved and novel products, create a stress on the basic sciences
of Physics and Chemistry. This is a result of trying to mantain a
balance in the triade ST\&I (Science, Technology and Innovation)
[1]. Particular questions involve, for example, the dissipation of
energy and heat transport in devices under high-levels of
excitation, namely, working in far-removed-from equilibrium
conditions and eventually involving ultrafast relaxation and
transport processes, as well as spatial motion in nanometric scales.
Another important aspect is the one of fluids under flow present in
certain production processes (e.g., in food engineering,
petrochemistry, etc...) whose performance depends on their
hydrodynamic properties [2]. Moreover it can be mentioned the
question of figure of merit in thermoelectric devices, that is,
relating currents of charges and of heat, particularly in the
nanometric scale [3].

It has been noticed that one of the complicated problems of the
nonequilibrium theory of transport processes in dense gases and
liquids is the fact that their kinetics and hydrodynamics are
intimately coupled, and must be treated simultaneously (e.g., see
Refs. [4-6]). Along the last decades Hydrodynamics has been
extensively treated resorting to the so-called Nonequilibrium
Molecular Dynamics (NMD for short). NMD is a computational method
created for modelling physical systems at the microscopic level,
being a good technique to study the molecular behavior of several
physical processes [7,8]. On the other hand, another very
satisfactory approach to deal with hydrodynamics within an ample
scope of nonequilibrium conditions consists in the kinetic theory
based on the Non-Equilibrium Statistical Ensemble Formalism (NESEF
for short) [9-14]. NESEF is a powerful formalism that provides an
elegant, practical, and physically clear picture for describing
irreversible processes, as for example in semiconductors far-from
equilibrium [15-17]. NESEF provides a way to go beyond standard (or
classical) Onsagerian hydrodynamics which involves restrictions,
namely, local equilibrium; linear relations between fluxes and
thermodynamic forces (meaning weak amplitudes in the motion) with
Onsager's symmetry laws holding; near homogeneous and static
movement (i.e., involving only smooth variation in space and time);
and weak and rapidly regressing fluctuations [18,19]. Hence, more
advanced approaches are required to lift these restrictions. In
phenomenological theories this corresponds to go from classical
irreversible thermodynamics to extended irreversible thermodynamics
[20-22]. This is what has been called \emph{generalized
hydrodynamics}, a question extensively debated for decades by the
Statistical Mechanics community. Several approaches have been used,
and a description can be consulted in Chapter 6 of the classical
book on the subject by Boon and Yip [23]. Introduction of nonlocal
effects for describing motions with influence of ever decreasing
wavelengths, going towards the very short limit, has been done in
terms of expansions in increasing powers of the wavenumber, which
consists in what is sometimes referred to as ``Higher-Order
Generalized Hydrodynamics" (HOGH for short), also dubbed Mesoscopic
Hydro-Thermodynamics (MHT for short) [24].

Within the scope of Mesoscopic Hydro-Thermodynamics we consider here
the question of transport of charge and of heat in n-doped polar
semiconductors in the presence of electric fields. The hierarchy of
equations of evolution for the density and energy density of
carriers and of energy density of phonons and together with those
for their fluxes of all orders, are obtained in the framework of the
nonlinear quantum kinetic theory that is based on NESEF
[11-13,25,26]. The electrical and thermal conductivities in such
nonequilibrium thermodynamic state and within a MHT of order 1, that
is, a description reduced to include the densities and their first
fluxes, are derived, and the influence of the order of the HOGH
(contracted description in terms of the densities and a reduced
number of higher-order fluxes) and of the sample size (macro to
nano) are discussed. The so-called Maxwell times [27,28] are
characterized and analyzed, with some numerical calculations and
figures being presented. Maxwell times are of fundamental relevance
for establishing the order of the contracted description of MHT to
be used, and of large influence on the behavior of transport
properties at short nanoscales.

\section{Theoretical Background}

The construction of a Mesoscopic Hydro-Thermodynamics for the
description of the movement of matter and energy in fluids under
nonequilibrium thermodynamic conditions and at the classical
mechanical level based on a generalized moments approach method to
the solution of a NESEF-based generalized Boltzmann equation [29],
is described elsewhere [30,31].

We consider here MHT at the quantum mechanical level, for dealing
with a system of carriers and phonons in n-doped polar
semiconductors in the presence of electric fields (up to 100 kV/cm)
which drive the system away from equilibrium. Moreover, the system
is taken to be in contact with an external thermostat at temperature
$T_{0}$.

The system is characterized at the microscopic level by the Hamiltonian
%
\begin{equation}
\hat{H} = \hat{H}_{e} + \hat{H}_{p} + \hat{H}_{ee} + \hat{H}_{ep} +
\hat{H}_{an} + \hat{H}_{e\mathcal{E}} + \hat{H}_{pR} \, ,
\end{equation}
consisting of the Hamiltonians of the free electrons and free
phonons, respectively
%
\begin{equation}
\hat{H}_{e} = \sum \limits_{\mathbf{k}} \epsilon
_{\mathbf{k}}c_{\mathbf{k}}^{\dag }c_{\mathbf{k}} \, ,
\end{equation}
%
\begin{equation}
\hat{H}_{p} = \sum \limits_{\mathbf{q\gamma}} \hbar
\omega_{\mathbf{q\gamma}}(b_{\mathbf{q\gamma}}^{\dag
}b_{\mathbf{q\gamma}} + 1/2) \, ,
\end{equation}
where $\epsilon_{\mathbf{k}}$ is the electrons' conduction-band
energy (spin index has been ignored), $\omega_{\mathbf{q} \gamma}$
is the phonon frequency dispersion relation with $\gamma$ indicating
the branch \textsc{lo}, \textsc{to}, \textsc{la}, \textsc{ta}, and
$\mathbf{k}$ and $\mathbf{q}$ are wave vectors running over the
Brillouin zone. The electron-electron interaction is
%
\begin{equation}
\hat{H}_{ee} = \sum \limits_{\substack{
\mathbf{k}_{1}\mathbf{k}_{2}\mathbf{k }_{3}\mathbf{k}_{4}  \\
(\mathbf{k}_{1}+\mathbf{k}_{2} = \mathbf{k}_{3} + \mathbf{k}_{4})}}
V(\mathbf{k}_{1}\mathbf{k}_{2}\mathbf{k}_{3}\mathbf{k}_{4})
c_{\mathbf{k}_{1}}^{\dag} c_{\mathbf{k}_{2}}^{\dag}
c_{\mathbf{k}_{3}} c_{\mathbf{k}_{4}} \, ,
\end{equation}
and for the electron-phonon interaction we have
%
\begin{equation}
\hat{H}_{ep} = \sum \limits_{\mathbf{kq}}
\mathcal{C}^{\alpha}_{\mathbf{kq}\gamma} b_{ \mathbf{q} \gamma}
c_{\mathbf{k}+\mathbf{q}}^{\dag }c_{\mathbf{k}} + \mathrm{ H.c.} \,
.
\end{equation}
where as noticed $\gamma$ indicates the phonon branch, and $\alpha$
the type of interaction (deformation potential, Fr\"{o}hlich-polar
with \textsc{lo} phonons, piezoelectric with \textsc{la} phonons)
with coupling strength $\mathcal{C}$. The electron-electric field
interaction is given by
%
\begin{equation}
\hat{H}_{e\mathcal{E}} = -e\mathbf{E}\cdot
\sum\limits_{j=1}^{N}\mathbf{r} _{j}=-ie\mathcal{E} \sum
\limits_{\mathbf{k^{\prime}Q}^{\prime}} \left[ \frac{
\partial \delta (\mathbf{Q}^{\prime})}{\partial \mathbf{Q}^{\prime}}\right]
c_{\mathbf{k^{\prime}}+\mathbf{Q^{\prime}}}^{\dag}c_{\mathbf{k^{\prime}}} \,
,
\end{equation}
with an electric field $\mathbf{E}$ of intensity $\mathcal{E}$
applied in, say, $z$-direction and we write for the anharmonic
interaction
%
\begin{equation}
\hat{H}_{an} = \sum \limits_{\substack{ \mathbf{kq}  \\ \gamma
\gamma^{\prime} \gamma^{\prime \prime}}} M_{\mathbf{kq}\gamma}
b_{\mathbf{q} \gamma} b_{\mathbf{k}+\mathbf{q}
\gamma^{\prime}}^{\dag} b_{-\mathbf{k} \gamma^{\prime
\prime}}^{\dag} + \mathrm{H.c.} \, ,
\end{equation}
(where we have neglected nonlinear contributions), with
$M_{\mathbf{k},\mathbf{q} \gamma}$ accounting for the coupling
strength.

For the description of the macroscopic nonequilibrium thermodynamic
state we resort, as already noticed, to the use of NESEF. The
statistical approach NESEF requires first of all to specify the
basic dynamical variables used to characterized the non-equilibrium
ensemble [9,11-13,25,32]. A priori, when the system is driven away
from equilibrium, it is necessary to include all observables of the
system, which leads to the introduction of many-particle dynamical
operators [33,34], in the present case of single electrons in Bloch
conduction band and single phonons, it suffices to introduce only
the single particle dynamical operator, namely
%
\begin{equation}
\hat{n}_{\mathbf{k}} = c_{\mathbf{k}}^{\dag }c_{\mathbf{k}} \qquad ;
\qquad \hat{ n}_{\mathbf{kQ}} = c_{\mathbf{k}+\mathbf{Q}/2}^{\dag}
c_{\mathbf{k}-\mathbf{Q}/2} \, ,
\end{equation}
with $\mathbf{Q}\neq 0$, and spin index ignored, for the carriers, and
%
\begin{equation}
\hat{\nu}_{\mathbf{q}\gamma} = b_{\mathbf{q}\gamma}^{\dag
}b_{\mathbf{q} \gamma} \qquad ; \qquad \hat{\nu}_{\mathbf{qQ}\gamma}
= b_{\mathbf{q}+\frac{1}{2}\mathbf{Q}\gamma}^{\dag}
b_{\mathbf{q}-\frac{1}{2}\mathbf{Q}\gamma} \, ,
\end{equation}
with $\mathbf{Q}\neq 0$ for the phonons.

Dynamical operators of order two and higher (in the BBGKY hierarchy
[34]) do not contribute because correlations and higher-order
variances are absent in the mean-field approximation for the
carriers and the harmonic approximation for the lattice vibrations.
Moreover, since phonons are bosons, it would be necessary also to
include the annihilation and creation operators
$\hat{b}_{\mathbf{q}}$ and $\hat{b}_{\mathbf{q}}^{\dag}$ because
their eingenstates are the coherent states [35], and also the pair
operators $\hat{b}_{\mathbf{k}}\hat{b}_{\mathbf{k}^{\prime}}$,
$\hat{b}_{\mathbf{k}}^{\dag}\hat{b}_{\mathbf{k}^{\prime}}^{\dag}$
because the number of quasi-particles is not fixed [36]. However, we
disregard them because are of no relevance for the problem at hands.
In Appendix A we describe the corresponding non-equilibrium
statistical operator (Cf. Eqs. (A1) and (A2)).

Operators $\hat{n}_{\mathbf{k}}$ and $\hat{\nu}_{\mathbf{q}\gamma }$
correspond to the occupation number operator describing a homogeneous
population and those with $\mathbf{Q}\neq 0$, account for changes in space
of the non-equilibrium distribution functions.

The average, over the non-equilibrium ensemble, of the microdynamical
variables in the sets of Eqs. (8) and (9) provide the variables which
characterize the non-equilibrium macroscopic state of the system. Let us
call them
%
\begin{equation}
\left\{ n_{\mathbf{k}}(t),n_{\mathbf{kQ}}(t),\nu_{\mathbf{q\gamma
}}(t),\nu_{\mathbf{qQ\gamma}}(t)\right\} \, ,
\end{equation}
where $\mathbf{Q}\neq 0$ and $n_{\mathbf{k}} (t) = \mathrm{Tr}
\left\{ \hat{n}_{\mathbf{k}} \varrho_{\varepsilon}(t)\right\}$,
etc..., that is, the average over the non-equilibrium ensemble
according to the formalism in Appendix A where we have introduced
the non-equilibrium thermodynamic state variables said conjugated to
those above, namely [cf. Eq. (A.3)]
%
\begin{equation}
\left\{ F_{\mathbf{k}}(t),F_{\mathbf{kQ}}(t),\varphi
_{\mathbf{q}\gamma}(t),\varphi_{\mathbf{qQ}\gamma}(t)\right\} \, .
\end{equation}

Going over to direct space we introduce the space and
crystal-momentum dependent distribution functions
%
\begin{equation}
\frac{1}{V_{cel}}\sum\limits_{\mathbf{Q}}n_{\mathbf{kQ}}(t)e^{i\mathbf{Q}
\cdot \mathbf{r}} = f_{\mathbf{k}}(\mathbf{r},t)\,,
\end{equation}
%
\begin{equation}
\frac{1}{V_{cel}} \sum\limits_{\mathbf{Q}}\nu
_{\mathbf{qQ}}(t)e^{i\mathbf{Q} \cdot \mathbf{r}} = \nu
_{\mathbf{q}}(\mathbf{r},t) \, ,
\end{equation}
where $V_{cel}$ is the volume of the unit cell, and the phonon
branch under index $\gamma$ is being implicit from here on.

In terms of this microscopic (quantum mechanical) and macroscopic
(nonequilibrium thermodynamic) description of the system, we proceed
to present the evolution equations of the basic variables in the set
of Eq. (10).

\section{Evolution of the Nonequilibrium Thermodynamic State}

Calling, in a compact and generic form, $\hat{P}_{j}$ and $Q_{j}(t)$
the dynamical variables and the corresponding thermodynamic
variables in the sets of Eqs. (8) and (9) and (10) respectively, the
evolution equations for the variables $Q_{j}(t)$ describing the
evolution of the nonequilibrium thermodynamic state of the system
are
%
\begin{equation}
\frac{d}{dt}Q_{j}(t) = \frac{d}{dt} \mathrm{Tr} \{\hat{P}_{j}
\varrho_{\varepsilon}(t) \times \varrho_{B} \} = \mathrm{Tr}\left\{
\frac{1}{i\hbar }[\hat{P}_{j},\hat{H}] \varrho_{\varepsilon}(t)
\times \varrho_{B} \right\} \, ,
\end{equation}
that is, the average over the nonequilibrium ensemble, characterized
by the statistical operator $\varrho _{\varepsilon }(t)$ of Appendix
A (Cf. Eqs. (A1) to (A3)), of the Heisenberg equation for the
corresponding dynamical variable $\hat{P}_{j}$; $\varrho_{B}$ is the
distribution of the surrounding medium assumed in equilibrium at
temperature $T_{0}$.

Direct calculation of the right-hand-side in Eq. (14) is extremely
difficult and then it is necessary to resort to the introduction of
a more practical non-linear kinetic theory [11,13,25,26] briefly
described in Appendix B, which is applied using an approximation
consisting in retaining only the collision integrals of second order
in the interaction strengths (Markovian approximation [26,29,32]).
In reciprocal space it follows that (see Appendix B)
%
\begin{equation}
\frac{d}{dt}n_{\mathbf{kQ}}(t) = \frac{1}{\hbar}(i \mathbf{Q} \cdot
\nabla_{\mathbf{k}} \epsilon_{\mathbf{k}})n_{\mathbf{kQ}}(t) +
J_{\mathbf{kQ}}^{ee}(t) + J_{\mathbf{kQ}}^{ep}(t) +
J_{\mathbf{kQ}}^{e\mathcal{E}}(t) + J_{\mathbf{kQ}}^{\nabla T}(t) \,
,
\end{equation}
%
\begin{equation}
\frac{d}{dt} \nu_{\mathbf{qQ}}(t) = (iQ\cdot \nabla
_{\mathbf{q}}\omega_{\mathbf{q}}) \nu_{\mathbf{qQ}}(t) +
J_{\mathbf{qQ}}^{an}(t) + J_{\mathbf{qQ}}^{pR}(t) +
J_{\mathbf{qQ}}^{\mathrm{ext.}}(t) \, ,
\end{equation}
where the seven collision integrals $J$'s are given in generic form
in Eqs. (B.3) to (B.6) in Appendix B. In direct space, after using
Eqs. (12) and (13) and for the different $J$'s that
%
\begin{equation}
J_{\mathbf{k}}(\mathbf{r},t) =
\frac{1}{V_{cel}}\sum\limits_{\mathbf{Q}}J_{
\mathbf{kQ}}(t)e^{i\mathbf{Q}\cdot \mathbf{r}} \, \quad ; \quad
J_{\mathbf{q}}(\mathbf{r},t) =
\frac{1}{V_{cel}}\sum\limits_{\mathbf{Q}}J_{\mathbf{qQ}}(t)e^{i
\mathbf{Q}\cdot \mathbf{r}} \, ,
\end{equation}
we do have that
%
\begin{equation}
\frac{\partial}{\partial t}f_{\mathbf{k}}(\mathbf{r},t) +
\frac{1}{\hbar} \nabla_{\mathbf{k}} \epsilon_{\mathbf{k}} \cdot
\nabla_{\mathbf{r}}f_{\mathbf{k}}(\mathbf{r},t) =
J_{\mathbf{k}}^{electrons}(\mathbf{r},t) \, ,
\end{equation}
%
\begin{equation}
\frac{\partial}{\partial t} \nu_{\mathbf{q}_{\gamma}}(\mathbf{r},t)
+ \nabla_{\mathbf{q}_{\gamma}} \omega_{\mathbf{q}_{\gamma}} \cdot
\nabla_{\mathbf{r}} \nu _{\mathbf{q}_{\gamma}}(\mathbf{r},t) =
J_{\mathbf{q}_{\gamma}}^{phonons}(\mathbf{r},t) \, ,
\end{equation}
where:
%
\begin{equation}
J_{\mathbf{k}}^{electrons}(\mathbf{r},t) =
J_{\mathbf{k}}^{e\text{-}p}(\mathbf{r},t) +
J_{\mathbf{k}}^{e\text{-}e}(\mathbf{r},t) +
J_{\mathbf{k}}^{e\text{-}s}(\mathbf{r},t) \, ,
\end{equation}
%
\begin{equation}
J_{\mathbf{q}_{\gamma}}^{phonons}(\mathbf{r},t) =
J_{\mathbf{q}_{\gamma}}^{p\text{-}e}(\mathbf{r},t) +
J_{\mathbf{q}_{\gamma}}^{an.}(\mathbf{r},t) +
J_{\mathbf{q}_{\gamma}}^{p\text{-}s}(\mathbf{r},t) \, ,
\end{equation}
where $J_{\mathbf{k}}^{e\text{-}p}(\mathbf{r},t)$ accounts for the
effect of the electron-phonon interaction,
$J_{\mathbf{k}}^{e\text{-}e}(\mathbf{r},t)$ of the internal
interaction (electron-electron),
$J_{\mathbf{k}}^{e\text{-}s}(\mathbf{r},t)$ the electron-sources
interactions, $J_{\mathbf{q}_{\gamma}}^{p\text{-}e}(\mathbf{r},t)$
of the phonon-electron interaction,
$J_{\mathbf{q}_{\gamma}}^{an.}(\mathbf{r},t)$ of the anharmonic
interaction and $J_{\mathbf{q}_{\gamma}}^{p\text{-}s}(\mathbf{r},t)$
of the phonon-sources interactions.

We consider now the Mesoscopic Hydro-Thermodynamic of the system,
which consists into introducing the densities of (quasi)particles
and of the energy and their fluxes of all order, namely, for the
electrons,
%
\begin{equation}
\{n_{e}(\mathbf{r},t),\mathbf{I}_{n_{e}}(\mathbf{r},t), \ldots
I_{n_{e}}^{[\ell]}(\mathbf{r},t) \ldots \} \, ,
\end{equation}
which we call MHT-carriers' family $n$, and
%
\begin{equation}
\{h_{e}(\mathbf{r},t),\mathbf{I}_{h_{e}}(\mathbf{r},t), \ldots
I_{h_{e}}^{[\ell ]}(\mathbf{r},t) \ldots \} \, ,
\end{equation}
the MHT-carriers' family $h$, where
%
\begin{equation}
n_{e}(\mathbf{r},t) =
\sum\limits_{\mathbf{k}}f_{\mathbf{k}}(\mathbf{r},t) \, ,
\end{equation}
%
\begin{equation}
\mathbf{I}_{n_{e}}(\mathbf{r},t) = \sum\limits_{\mathbf{k}}\nabla
_{\mathbf{k}}\epsilon_{\mathbf{k}}f_{\mathbf{k}}(\mathbf{r},t) \, ,
\end{equation}
%
\begin{equation}
I_{n_{e}}^{[\ell]}(\mathbf{r},t) = \sum \limits_{\mathbf{k}}
u_{e,\mathbf{k}}^{[\ell]}f_{\mathbf{k}}(\mathbf{r},t) \, ,
\end{equation}
with $\ell =2,3,...$ and
%
\begin{equation}
u_{\epsilon \mathbf{k}}^{[\ell]} = \frac{1}{\hbar^{\ell}}
[\nabla_{\mathbf{k}} \epsilon_{\mathbf{k}} \colon \ldots \ell \, \,
\text{times} \ldots  \colon \nabla_{\mathbf{k}}
\epsilon_{\mathbf{k}}] \, ,
\end{equation}
is a rank-$\ell$ tensor involving $\ell$-times the tensorial
internal product of the group velocity
$(1/\hbar)\nabla_{\mathbf{k}}\epsilon_{\mathbf{k}}$ (in a effective
mass approximation $\epsilon_{\mathbf{k}} =
\hbar^{2}k^{2}/2m_{e}^{\ast}$, and then $\nabla_{\mathbf{k}}
\epsilon_{\mathbf{k}} = \hbar \mathbf{k} /m_{e}^{\ast}$;
$m_{e}^{\ast}$ is the effective mass of the electrons at the center
of the conduction Bloch band in polar semiconductors), and
%
\begin{equation}
h_{e}(\mathbf{r},t) = \sum\limits_{\mathbf{k}}
\epsilon_{\mathbf{k}}f_{\mathbf{k}}(\mathbf{r},t) \, ,
\end{equation}
%
\begin{equation}
\mathbf{I}_{h_{e}}(\mathbf{r},t) = \sum\limits_{\mathbf{k}}\epsilon
_{\mathbf{k}}\nabla_{\mathbf{k}}\epsilon
_{\mathbf{k}}f_{\mathbf{k}}(\mathbf{r},t) \, ,
\end{equation}
%
\begin{equation}
I_{h_{e}}^{[\ell ]}(\mathbf{r},t) = \sum\limits_{\mathbf{k}}\epsilon
_{\mathbf{k}}u_{e,\mathbf{k}}^{[\ell]}f_{\mathbf{k}}(\mathbf{r},t)
\, .
\end{equation}

On the other hand, for the phonons we do have
%
\begin{equation}
\{n_{p}(\mathbf{r},t), \mathbf{I}_{n_{p}}(\mathbf{r},t), \ldots ,
I^{[\ell]}_{n_{p}}(\mathbf{r},t) ,\ldots \} \, ,
\end{equation}
the MHT-phonons' family $n$, and
%
\begin{equation}
\{h_{p}(\mathbf{r},t), \mathbf{I}_{h_{p}}(\mathbf{r},t), \ldots ,
I^{[\ell]}_{h_{p}}(\mathbf{r},t) ,\ldots \} \, ,
\end{equation}
the MHT-phonons' family $h$, where
%
\begin{equation}
n_{p}(\mathbf{r},t) = \sum\limits_{\mathbf{q}}\nu
_{\mathbf{q}}(\mathbf{r},t) \, ,
\end{equation}
%
\begin{equation}
\mathbf{I}_{p}(\mathbf{r},t) = \sum \limits_{\mathbf{q}}\hbar \omega
_{\mathbf{q}}\nabla_{\mathbf{q}} \omega_{\mathbf{q}} \,
\nu_{\mathbf{q}}(\mathbf{r},t) \, ,
\end{equation}
%
\begin{equation}
I_{p}^{[\ell ]}(\mathbf{r},t) = \sum \limits_{\mathbf{q}} \hbar
\omega _{\mathbf{q}}u_{ph,\mathbf{q}}^{[\ell]} \,
\nu_{\mathbf{q}}(\mathbf{r},t) \, ,
\end{equation}
for the MHT-phonons' $n$-family, and
%
\begin{equation}
h_{p}(\mathbf{r},t) = \sum \limits_{\mathbf{q}} \hbar \omega
_{\mathbf{q}} \, \nu_{\mathbf{q}}(\mathbf{r},t) \, ,
\end{equation}
%
\begin{equation}
\mathbf{I}_{h_{p}}(\mathbf{r},t) = \sum \limits_{\mathbf{q}}\hbar
\omega_{ \mathbf{q}}\nabla _{\mathbf{q}}\omega_{\mathbf{q}} \,
\nu_{\mathbf{q}}(\mathbf{r},t)\,,
\end{equation}
%
\begin{equation}
I_{h_{p}}^{[\ell ]}(\mathbf{r},t) = \sum\limits_{\mathbf{q}}\hbar
\omega_{\mathbf{q}}u_{ph,\mathbf{q}}^{[\ell ]} \,
\nu_{\mathbf{q}}(\mathbf{r},t) \, ,
\end{equation}
for the MHT-phonons' $h$-family, and where
%
\begin{equation}
u_{ph,\mathbf{q}}^{[\ell]} = [\nabla_{\mathbf{q}}\omega
_{\mathbf{q}}  \colon \ldots \ell \, \, \text{times} \ldots  \colon
\nabla_{\mathbf{q}}\omega_{\mathbf{q}}] \, ,
\end{equation}
and we recall that the phonon branch index $\gamma$ is implicit;
$\nabla_{\mathbf{q}}\omega_{\mathbf{q}}$ is the group velocity of
phonons in mode $\mathbf{q}$.

The evolution equations which describe the hydrodynamic motion in
MHT are:
%
\begin{equation}
\frac{\partial }{\partial t}I_{n_{e}}^{[\ell ]}(\mathbf{r},t) = \sum
\limits_{\mathbf{k}} u_{e}^{[\ell]}(\mathbf{k})\frac{\partial
}{\partial t}f_{ \mathbf{k}}(\mathbf{r},t)\,,
\end{equation}
%
\begin{equation}
\frac{\partial }{\partial t}I_{h_{e}}^{[\ell]}(\mathbf{r},t) = \sum
\limits_{\mathbf{k}} \epsilon_{\mathbf{k}}
u_{e}^{[\ell]}(\mathbf{k}) \frac{\partial}{\partial t}
f_{\mathbf{k}}(\mathbf{r},t) \, ,
\end{equation}
%
\begin{equation}
\frac{\partial}{\partial t}I_{n_{p}}^{[\ell ]}(\mathbf{r},t) = \sum
\limits_{\mathbf{q}} u_{ph}^{[\ell]}(\mathbf{q}) \frac{\partial
}{\partial t}\nu_{ \mathbf{q}}(\mathbf{r},t) \, ,
\end{equation}
%
\begin{equation}
\frac{\partial}{\partial t}I_{h_{p}}^{[\ell ]}(\mathbf{r},t) = \sum
\limits_{\mathbf{q}} \hbar \omega_{\mathbf{q}}
u_{ph}^{[\ell]}(\mathbf{q}) \frac{\partial }{\partial t}\nu_{
\mathbf{q}}(\mathbf{r},t) \, ,
\end{equation}
where, we recall, $\ell =0$ stands for the densities, $\ell =1$ for
their first (vectorial) fluxes, $\ell = 2,3,\ldots$ for the
higher-order tensorial fluxes.

This set of equations is practically intractable, requiring to look
in each case on how to find the best description using the smallest
possible number of variables. In other words to introduce an
appropriate -- for each case -- contraction of description:
\emph{this contraction implies in retaining the information
considered as relevant for the problem in hands, and to disregard
irrelevant information} [37].

Elsewhere [38] it has been discussed the question of the contraction
of description (reduction of the dimensions of the nonequilibrium
thermodynamic space of states), where a criterion for justifying the
different levels of contraction is derived: It depends on the range
of wavelengths and frequencies which are relevant for the
characterization, in terms of normal modes, of the
hydro-thermodynamic motion in the nonequilibrium open system. It can
be shown that the truncation criterion \emph{rests on the
characteristics of the hydrodynamic motion that develops under the
given experimental procedure}.

Since inclusion of higher and higher-order fluxes implies in
describing a motion involving increasing Knudsen numbers per
hydrodynamic mode (that is, governed by smaller and smaller
wavelengths -- larger and larger wavenumbers -- accompanied by
higher and higher frequencies). In a qualitative manner, we can say
that, as a general ``thumb rule", the criterion indicates that
\emph{a more and more restricted contraction can be used when larger
and larger are the prevalent wavelengths in the motion}. Therefore,
in simpler words, when the motion becomes more and more smooth in
space and time, the more reduced can be the dimension of the basic
macrovariables space to be used for the description of the
nonequilibrium thermodynamic state of the system. It can be
conjectured a general contraction criterion, namely, a contraction
of order $r$ (meaning keeping the densities and their fluxes up to
order $r$), once we can show that in the spectrum of wavelengths,
which characterizes the motion, predominate those larger than a
``frontier" one, $\lambda^{2}_{(r,r+1)} =
v^{2}\theta_{r}\theta_{r+1}$ where $v$ is of the order of the
thermal velocity and $\theta_{r}$ and $\theta_{r+1}$ the
corresponding \emph{Maxwell times}, see next, associated to the $r$
and $r+1$ order fluxes. We shall try next to illustrate the matter
using a contraction of order 1, that is, a first-order extension of
standard Onsagerian hydrodynamics.

\section{MHT of Order 1 of Carriers and Phonons}

We consider the contracted MHT of order 1 (that is keeping only the
densities and their first fluxes, implying in smooth movement in
space and slow in time, but beyond the range in standard
hydrodynamics) in the already described system of carriers and
phonons in a n-doped polar semiconductor in the presence of an
electric field. Hence, the basic sets of dynamical variables are
%
\begin{equation}
\left\{
\hat{n}_{e}(\mathbf{r}),\mathbf{\hat{I}}_{n_{e}}(\mathbf{r}),\hat{h}
_{e}(\mathbf{r}),\mathbf{\hat{I}}_{h_{e}}(\mathbf{r})\right\}
\end{equation}
for the carriers, and
%
\begin{equation}
\left\{
\hat{n}_{p}(\mathbf{r}),\mathbf{\hat{I}}_{n_{p}}(\mathbf{r}),\hat{h}
_{p}(\mathbf{r}),\mathbf{\hat{I}}_{h_{p}}(\mathbf{r})\right\} \,,
\end{equation}
for the phonons, or in reciprocal space
%
\begin{equation}
\left\{
\hat{n}_{e}(\mathbf{Q}),\mathbf{\hat{I}}_{n_{e}}(\mathbf{Q}),\hat{h}
_{e}(\mathbf{Q}),\mathbf{\hat{I}}_{h_{e}}(\mathbf{Q})\right\} \, ,
\end{equation}
and
%
\begin{equation}
\left\{
\hat{n}_{p}(\mathbf{Q}),\mathbf{\hat{I}}_{n_{p}}(\mathbf{Q}),\hat{h}
_{p}(\mathbf{Q}),\mathbf{\hat{I}}_{h_{p}}(\mathbf{Q})\right\} \, .
\end{equation}

The associated auxiliary statistical operator $\bar{\varrho}$ (see
Appendix A) is then
%
\begin{eqnarray}
\bar{\varrho}(t,0) &=&\exp \Big{\{}-\phi (t)  \notag \\
&&-\sum\limits_{\mathbf{Q}}[F_{n_{e}}(\mathbf{Q},t)\hat{n}_{e}(\mathbf{Q})
+ \mathbf{F}_{n_{e}}(\mathbf{Q},t)\cdot
\mathbf{\hat{I}}_{n_{e}}(\mathbf{Q},t) + \notag \\
&& +
F_{h_{e}}(\mathbf{Q},t)\hat{h}_{e}(\mathbf{Q},t)+\mathbf{F}_{h_{e}}(
\mathbf{Q},t)\cdot \mathbf{\hat{I}}_{h_{e}}(\mathbf{Q},t) +  \notag \\
&&+F_{n_{p}}(\mathbf{Q},t)\hat{n}_{p}(\mathbf{Q})+\mathbf{F}_{n_{p}}(\mathbf{
Q},t)\cdot \mathbf{\hat{I}}_{n_{p}}(\mathbf{Q}) +  \notag \\
&&+F_{h_{p}}(\mathbf{Q},t)\hat{h}_{p}(\mathbf{Q}) +
\mathbf{F}_{h_{p}}(\mathbf{ Q},t)\cdot
\mathbf{\hat{I}}_{h_{p}}(\mathbf{Q})]\Big{\}} \; ,
\end{eqnarray}
introducing the set of nonequilibrium thermodynamic variables
%
\begin{equation}
\left\{
F_{n_{e}}(\mathbf{Q},t),\mathbf{F}_{n_{e}}(\mathbf{Q},t),F_{h_{e}}(
\mathbf{Q},t),\mathbf{F}_{h_{e}}(\mathbf{Q},t),F_{n_{p}}(\mathbf{Q},t),
\mathbf{F}_{n_{p}}(\mathbf{Q},t),F_{h_{p}}(\mathbf{Q},t),\mathbf{F}_{h_{p}}(\mathbf{Q},t)
\right\} \, ,
\end{equation}
where $\mathbf{Q}=0$ refers to the homogeneous (or global)
state-thermodynamic variables, and $\mathbf{Q}\neq 0$ to the
inhomogeneous contributions. As usually done, we write
%
\begin{equation}
F_{h_{e}}(0,t) \equiv -\beta _{e}(t) \equiv
-[k_{B}T_{e}^{\ast}(t)]^{-1},
\end{equation}
introducing the carriers quasitemperature $T_{e}^{\ast}(t)$,
%
\begin{equation}
\mathbf{F}_{n_{e}}(0,t) \equiv \beta _{e}(t) \mathbf{v}_{e}(t) \, ,
\end{equation}
with $\mathbf{v}_{e}(t)$ being the drift velocity, and
%
\begin{equation}
F_{n_{e}}(0,t) \equiv \beta _{e}(t) \mu _{e}^{\ast }(t) \, ,
\end{equation}
introducing the quasi-chemical potential $\mu_{e}^{\ast}$.

On the other hand, for the phonons we do have that
$F_{n_{p}}(0,t)=0$ (number of phonons not conserved),
$\mathbf{F}_{n_{p}}(0,t)$ (no close current circuit present), and we
write
%
\begin{equation}
F_{h_{p}}(0,t) \widehat{h}_{p}(0) \equiv - \sum
\limits_{\mathbf{q\gamma}} \beta _{\mathbf{q\gamma}}(t) \hbar
\omega_{\mathbf{q\gamma}} b_{\mathbf{q\gamma}}^{\dag}
b_{\mathbf{q\gamma}} \, ,
\end{equation}
with
%
\begin{equation}
\beta_{\mathbf{q\gamma}}(t) \equiv \lbrack
k_{B}T_{\mathbf{q\gamma}}^{\ast}(t)]^{-1} ,
\end{equation}
introducing the phonon quasitemperature per mode in each
branch$T_{\mathbf{q\gamma}}^{\ast}(t)$.

The set of evolution equations, see Appendix B, for the electrons
results in that
%
\begin{equation}
\frac{d}{dt}n_{e}(\mathbf{Q},t) = i\mathbf{Q}\cdot
\mathbf{I}_{n_{e}}(\mathbf{ Q},t) + \Phi_{e}(\mathbf{Q},t) \, ,
\end{equation}
%
\begin{eqnarray}
\frac{d}{dt}\mathbf{I}_{n_{e}}(\mathbf{Q},t) &=& i\mathbf{Q}\cdot
\,I_{n_{e}}^{[2]}(\mathbf{Q},t)  \notag \\
&& - \sum \limits_{\mathbf{kq} \gamma \alpha}(\nabla
_{\mathbf{k}+\mathbf{q}} \epsilon_{\mathbf{k}+\mathbf{q}} -
\nabla_{\mathbf{k}} \epsilon_{\mathbf{k}}) A_{\mathbf{kq}_{\gamma}
\alpha}(t) (n_{e,\mathbf{kQ}}(t) - n_{e,\mathbf{k},-\mathbf{Q}}(t))  \notag \\
&&+e\mathbf{E}n_{e}(\mathbf{Q},t) \, ,
\end{eqnarray}
%
\begin{eqnarray}
\frac{d}{dt}h_{e}(\mathbf{Q},t) &=& i\mathbf{Q} \cdot
\mathbf{I}_{h_{e}}(\mathbf{Q},t)  \notag \\
&& - \sum \limits_{\mathbf{kq} \gamma \alpha}
(\epsilon_{\mathbf{k}+\mathbf{q}} - \epsilon_{\mathbf{k}}) A_{\mathbf{kq}_{\gamma} \alpha}(t)
(n_{e,\mathbf{kQ}}(t) + n_{e,\mathbf{k},-\mathbf{Q}}(t))  \notag \\
&& + e\frac{\mathbf{E}}{m_{e}^{\ast}}\cdot
\mathbf{I}_{e}(\mathbf{Q},t)n_{e}(\mathbf{Q},t) \, ,
\end{eqnarray}
%
\begin{eqnarray}
\frac{d}{dt}\mathbf{I}_{h_{e}}(\mathbf{Q},t) &=& i \mathbf{Q} \cdot
\mathbf{I}_{h_{e}}^{[2]}(\mathbf{Q},t)  \notag \\
&& - \sum \limits_{\mathbf{kq}\gamma \alpha} (\epsilon_{\mathbf{k}}
\nabla_{\mathbf{k}} \epsilon_{\mathbf{k}} -
\epsilon_{\mathbf{k}+\mathbf{q}} \nabla_{\mathbf{k} + \mathbf{q}}
\epsilon_{\mathbf{k}+\mathbf{q}}) A_{\mathbf{kq}_{\gamma} \alpha}(t)
n_{\mathbf{kQ}}(t) \notag \\
&& + \mathbf{I}_{n_{e}}^{[2]}(\mathbf{Q},t)\cdot e\mathbf{E} \, ,
\end{eqnarray}
where
%
\begin{eqnarray}
A_{\mathbf{kq\gamma ,\alpha}}(t) &=& \frac{2 \pi}{\hbar}
|\mathcal{C}_{\mathbf{kq\gamma } }^{\alpha }|^{2} \big\{
[(\nu_{\mathbf{q\gamma }}(t)+1) f_{\mathbf{k}+\mathbf{q}
}(t) + \nu _{\mathbf{q\gamma }}(t)(1-f_{\mathbf{k}+\mathbf{q}}(t))]  \notag \\
&&\times \delta (\epsilon_{\mathbf{k} + \mathbf{q}} -
\epsilon_{\mathbf{k}} - \hbar \omega_{\mathbf{q\gamma}})  \notag \\
&& + [(\nu_{\mathbf{q}\gamma}(t) + 1) (1 -
f_{\mathbf{k}+\mathbf{q}}(t))
+ \nu_{\mathbf{q}\gamma}(t) f_{\mathbf{k}+\mathbf{q}}(t)]  \notag \\
&& \times \delta (\epsilon_{\mathbf{k}+\mathbf{q}} -
\epsilon_{\mathbf{k}} + \hbar \omega_{\mathbf{q}\gamma}) \big\} \, ,
\end{eqnarray}
with the presence of the homogeneous distributions
%
\begin{equation}
f_{\mathbf{k}}(t) = \frac{1}{e^{\beta _{e}(t)[\epsilon_{\mathbf{k}}
- \mu_{e}^{\ast}(t) - \mathbf{v}_{e}(t) \cdot \mathbf{k}]} + 1} \, ,
\end{equation}
resembling a kind of shifted instantaneous Fermi-Dirac distribution,
which in the nondegenerate limit becomes
%
\begin{equation}
f_{\mathbf{k}}(t) = 4n \sqrt{\left( \frac{\pi \beta_{e}(t)
\hbar^{2}}{2m^{\ast}} \right)^{3}} \times e^{- \beta_{e}(t)(\hbar
\mathbf{k} - m^{\ast} \mathbf{v}(t))^{2}/2m^{\ast}} \, ,
\end{equation}
i.e., a shifted Maxwell-Boltzmann-like distribution where $n$ is the
density of carriers, and
%
\begin{equation}
\nu_{\mathbf{q\gamma}}(t) = \frac{1}{e^{\beta_{\mathbf{q\gamma}}(t)
\hbar \omega_{\mathbf{q\gamma}}} - 1} \, ,
\end{equation}
which has the form of a Bose-Einstein like distribution at zero
quasi-chemical potential. Equations (59) and (61) are a result of
the calculation of
\begin{equation*}
f_{\mathbf{k}}(t) = \mathrm{Tr} \{c_{\mathbf{k}}^{\dag}
c_{\mathbf{k}} \bar{\varrho}(t,0)\} \, , \quad \mathrm{and} \quad
\nu_{\mathbf{q\gamma}}(t) = \mathrm{Tr}
\{b_{\mathbf{q}\gamma}^{\dag} b_{\mathbf{q\gamma}}
\bar{\varrho}(t,0) \} \, ,
\end{equation*}
and the use of Eqs. (50), (51), (52) and (54).

The scattering integral $\Phi_{e}$ in Eq. (55) accounts for local
effects due to the presence of impurities, imperfections
(dislocations, stacking faults, etc.), the imperfections in the end
contacts, and geometry and boundary influences; the lateral walls
are rugous (of a fractal-on-average character [39]) leading to
inhomogeneous scattering of the carriers. The integration of
$\Phi_{e}(\mathbf{r},t)$ over the volume of the sample is null since
the total number of carriers is constant. We recall that
$\mathcal{C}_{\mathbf{kq}\gamma}^{\alpha}$ is the matrix element of
the electron-phonon interaction [cf. Eq. (5)]; it may be noticed
that in polar semiconductors Fr\"{o}hlich-polar interaction
($\gamma=$ \textsc{lo}, $\alpha =$ Fr\"{o}hlich interaction) is by
far the relevant one producing rates of change orders of magnitude
greater than those associated to the other interactions [40,41].
Moreover, we have neglected the contribution of the plasma states
via Coulomb interaction.

For the phonons we do have,
%
\begin{eqnarray}
\frac{d}{dt}n_{p}(\mathbf{Q},t) &=& i\mathbf{Q} \cdot
\mathbf{I}_{n_{p}}(\mathbf{Q},t) - \frac{1}{2} \sum
\limits_{\mathbf{q}\gamma \alpha} [
\Gamma_{\mathbf{q}+\mathbf{Q}/2,\gamma \alpha}^{e\text{-}p}(t) +
\Gamma_{\mathbf{q}-\mathbf{Q}/2,\gamma \alpha}^{e\text{-}p}(t) ]
\nu_{\mathbf{qQ\gamma}}(t)  \notag \\
&& - \frac{1}{2} \sum \limits_{\mathbf{q}\gamma \alpha} [
\Gamma_{\mathbf{q}+\mathbf{Q} /2,\gamma \alpha}^{an}(t) +
\Gamma_{\mathbf{q}-\mathbf{Q}/2,\gamma \alpha}^{an}(t) ]
\nu_{\mathbf{qQ\gamma}}(t) \, ,
\end{eqnarray}
%
\begin{eqnarray}
\frac{d}{dt}\mathbf{I}_{n_{p}}(\mathbf{Q},t) &=& i\mathbf{Q} \cdot
I_{n_{p}}^{[2]}(\mathbf{Q},t)  \notag \\
&& - \frac{1}{2} \sum \limits_{\mathbf{q}\gamma \alpha} [
\Gamma_{\mathbf{q}+\mathbf{Q} /2,\gamma \alpha}^{e\text{-}p}(t) +
\Gamma_{\mathbf{q}-\mathbf{Q}/2,\gamma \alpha}^{e\text{-}p}(t) ]
\nabla_{\mathbf{q}} \omega_{\mathbf{q\gamma}} \nu_{\mathbf{qQ\gamma}}(t)  \notag \\
&& - \frac{1}{2} \sum \limits_{\mathbf{q}\gamma \alpha} [
\Gamma_{\mathbf{q}+\mathbf{Q} /2,\gamma \alpha}^{an}(t) +
\Gamma_{\mathbf{q}-\mathbf{Q}/2,\gamma \alpha}^{an}(t) ]
\nabla_{\mathbf{q}} \omega_{\mathbf{q\gamma}}\nu_{\mathbf{qQ\gamma
}}(t)\,,
\end{eqnarray}
%
\begin{eqnarray}
\frac{d}{dt}h_{p}(\mathbf{Q},t) &=& i\mathbf{Q} \cdot
\mathbf{I}_{h_{p}}(\mathbf{Q},t)  \notag \\
&& - \frac{1}{2} \sum \limits_{\mathbf{q}\gamma \alpha} [
\Gamma_{\mathbf{q}+\mathbf{Q} /2,\gamma \alpha}^{e\text{-}p}(t) +
\Gamma _{\mathbf{q}-\mathbf{Q}/2,\gamma \alpha}^{e\text{-}p}(t) ]
\hbar \omega_{\mathbf{q\gamma}} \nu_{\mathbf{qQ\gamma}}(t)  \notag \\
&& - \frac{1}{2} \sum \limits_{\mathbf{q}} [
\Gamma_{\mathbf{q}+\mathbf{Q}/2,\gamma \alpha}(t) +
\Gamma_{\mathbf{q}-\mathbf{Q}/2,\gamma \alpha}(t) ] \hbar
\omega_{\mathbf{q\gamma}} \nu_{\mathbf{qQ\gamma}}(t) \, ,
\end{eqnarray}
%
\begin{eqnarray}
\frac{d}{dt}\mathbf{I}_{h_{p}}(\mathbf{Q},t) &=& i\mathbf{Q} \cdot
I_{h_{p}}^{[2]}(\mathbf{Q},t)  \notag \\
&& - \frac{1}{2} \sum \limits_{\mathbf{q}\gamma \alpha} \hbar
\omega_{\mathbf{q} \gamma} \nabla_{\mathbf{q}}
\omega_{\mathbf{q\gamma}} [ \Gamma_{\mathbf{q}+\mathbf{Q} /2,\gamma
\alpha}^{e\text{-}p}(t) +
\Gamma_{\mathbf{q}-\mathbf{Q}/2,\gamma \alpha}^{e\text{-}p}(t) ] \nu_{\mathbf{qQ} \gamma}(t)  \notag \\
&& - \frac{1}{2} \sum \limits_{\mathbf{q}\gamma \alpha} \hbar
\omega_{\mathbf{q\gamma}} \nabla_{\mathbf{q}}
\omega_{\mathbf{q\gamma}} [ \Gamma_{\mathbf{q}+\mathbf{ Q}/2,\gamma
\alpha}(t) + \Gamma_{\mathbf{q}-\mathbf{Q}/2,\gamma \alpha}(t) ]
\nu_{\mathbf{qQ} \gamma}(t) \, ,
\end{eqnarray}
where
%
\begin{eqnarray}
\Gamma_{\mathbf{q\gamma} \alpha}^{e\text{-}p}(t) &=& \sum
\limits_{\mathbf{k}^{\prime}}|
\mathcal{C}_{\mathbf{qk}^{\prime}\gamma}^{\alpha }|^{2} \big\{
f_{\mathbf{k}^{\prime}+\mathbf{q}}(t) [1 -
f_{\mathbf{k}^{\prime}}(t)] - f_{\mathbf{k}^{\prime}}(t)
[1-f_{\mathbf{k}^{\prime}+\mathbf{q}}(t)] \big\} \times  \notag \\
&& \delta (\epsilon_{\mathbf{k}^{\prime}} - \epsilon_{\mathbf{k}} -
\hbar \omega_{\mathbf{q\gamma}}) \, ,
\end{eqnarray}
with dimension of inverse of time, accounts for the rate of transfer
(energy and momentum) from the hot carriers, and
%
\begin{equation}
\Gamma_{\mathbf{q}\gamma \alpha }(t) = \frac{\pi}{\hbar ^{2}} \sum
\limits_{\mathbf{q}^{\prime}} |M_{\mathbf{q\gamma q}^{\prime} \gamma
^{\prime}}^{\alpha}|^{2} (1 + \nu_{\mathbf{q}^{\prime}
\gamma^{\prime}} + \nu_{\mathbf{q}+\mathbf{q}^{\prime} \gamma
})\delta (\omega _{\mathbf{q}+ \mathbf{q}^{\prime }\gamma } + \omega
_{\mathbf{q}^{\prime }\gamma ^{\prime}} - \omega_{\mathbf{q\gamma
}}) + \overline{\Gamma}_{\mathbf{q}}  \, ,
\end{equation}
with the first contribution on the right being the explicit
expression for the inverse of the relaxation time due to anharmonic
interactions, and $\overline{\Gamma}$ stands for, in a
Mathiessen-like rule, the sum of the reciprocals of the relaxation
times associated to the interaction with impurities, imperfections,
stacking faults, as well as effects of (rugous) boundary conditions
and contacts with other subsystems and sources.

Evidently, the set of Eqs. (63) to (66) is not a closed one, once
the right-hand sides are not given in terms of the proper basic
hydrodynamic variables. Hence, we must proceed to introduce a
closure condition, what is done resorting to Heims-Jaynes
perturbation procedure for averages [42]. This is described in
Appendix C, and in a first-order linear approach in Heims-Jaynes
procedure, the fundamental set of hydrodynamic equations in MHT of
order 1 are for the carriers
%
\begin{equation}
\frac{d}{dt}n_{e}(\mathbf{Q},t) = i\mathbf{Q}\cdot
\mathbf{I}_{n_{e}}(\mathbf{Q},t) + \Phi _{e}(\mathbf{Q},t) \, ,
\end{equation}
%
\begin{eqnarray}
\frac{d}{dt}\mathbf{I}_{n_{e}}(\mathbf{Q},t) &=&
B_{1e}^{[2]}i\mathbf{Q}n_{e}(
\mathbf{Q},t)+B_{2e}^{[2]}i\mathbf{Q}h_{e}\,(\mathbf{Q},t)-\theta
_{I_{n_{e}}}^{-1}\mathbf{I}_{n_{e}}(\mathbf{Q},t)  \notag \\
&&+b_{34e}\mathbf{I}_{h_{e}}(\mathbf{Q},t)+\frac{e\mathbf{E}}{m_{e}^{\ast
}} n_{e}(\mathbf{Q},t) \, ,
\end{eqnarray}
%
\begin{eqnarray}
\frac{d}{dt}h_{e}(\mathbf{Q},t) &=& i \mathbf{Q}\cdot
\mathbf{I}_{h_{e}}( \mathbf{Q},t) +
b_{21e}n_{e}(\mathbf{Q},t)-\theta
_{h_{e}}^{-1}h_{e}(\mathbf{Q},t)  \notag \\
&&+e\mathbf{E}\cdot \mathbf{I}_{n_{e}}(\mathbf{Q},t)n_{e} \, ,
\end{eqnarray}
%
\begin{eqnarray}
\frac{d}{dt}\mathbf{I}_{h_{e}}(\mathbf{Q},t)
&=&C_{1e}^{[2]}i\mathbf{Q}n_{e}(
\mathbf{Q},t)+C_{2e}^{[2]}i\mathbf{Q}h_{e}(\mathbf{Q},t)+b_{43e}\mathbf{I}
_{n_{e}}(\mathbf{Q},t)  \notag \\
&&-\theta _{I_{h_{e}}}^{-1}\mathbf{I}_{h_{e}}(\mathbf{Q}
,t)+B_{1e}^{[2]}n_{e}(\mathbf{Q},t)\mathbf{E}+B_{2e}^{[2]}h_{e}(\mathbf{Q},t)
\mathbf{E} \, ,
\end{eqnarray}
where
%
\begin{equation}
B_{1e}^{[2]}(t) = \sum \limits_{\mathbf{k}}[\nabla
_{\mathbf{k}}\epsilon_{ \mathbf{k}}\colon \nabla
_{\mathbf{k}}\epsilon_{\mathbf{k}}]b_{1e}(\mathbf{k },t) \, ,
\end{equation}
%
\begin{equation}
B_{2e}^{[2]}(t)=\sum\limits_{\mathbf{k}}\epsilon
_{\mathbf{k}}[\nabla _{ \mathbf{k}}\epsilon _{\mathbf{k}}\colon
\nabla_{\mathbf{k}}\epsilon_{ \mathbf{k}}]b_{2e}(\mathbf{k},t) \, ,
\end{equation}
%
\begin{equation}
C_{1e}^{[2]}(t) = \sum \limits_{\mathbf{k}} \epsilon _{\mathbf{k}}
[\nabla_{\mathbf{k}} \epsilon_{\mathbf{k}} \colon
\nabla_{\mathbf{k}} \epsilon_{\mathbf{k}}]b_{1e}(\mathbf{k},t) \, ,
\end{equation}
%
\begin{equation}
C_{2e}^{[2]}(t) = \sum \limits_{\mathbf{k}}[\nabla
_{\mathbf{k}}\epsilon _{ \mathbf{k}}\colon
\nabla_{\mathbf{k}}\epsilon _{\mathbf{k}}]b_{2e}(\mathbf{k },t) \, ,
\end{equation}
with
%
\begin{equation}
b_{1e}(\mathbf{k},t) = [\Delta_{12e}(t)]^{-1} [A_{22e}(t) - \epsilon
_{\mathbf{k}}A_{12e}(t)] f_{\mathbf{k}}(t) [1 - f_{\mathbf{k}}(t)]
\, ,
\end{equation}
%
\begin{equation}
b_{2e}(\mathbf{k},t) = [\Delta_{12e}(t)]^{-1} [A_{11e}(t)
\epsilon_{\mathbf{k}} - A_{21e}(t)] f_{\mathbf{k}}(t)[1 -
f_{\mathbf{k}}(t)] \, ,
\end{equation}
%
\begin{equation}
A_{11e}(t) = \sum_{\mathbf{k}} f_{\mathbf{k}}(t)[1 -
f_{\mathbf{k}}(t)] \, ,
\end{equation}
%
\begin{equation}
A_{12e}(t) = A_{21e}(t) = \sum_{\mathbf{k}} \epsilon_{\mathbf{k}}
f_{\mathbf{k}}(t)[1 - f_{\mathbf{k}}(t)] \, ,
\end{equation}
%
\begin{equation}
A_{22e}(t) = \sum_{\mathbf{q}}(\epsilon_{\mathbf{k}})^{2}
f_{\mathbf{k}}(t)[1 - f_{\mathbf{k}}(t)] \, ,
\end{equation}
%
\begin{equation}
\Delta _{12e}(t) = A_{11e}(t)A_{22e}(t)-A_{12e}(t)A_{21e}(t) \, .
\end{equation}
with $f_{\mathbf{k}}(t)$ of Eq. (60), and we recall that $[\ldots
\colon \ldots]$ stands for inner tensorial product of two vectors
producing a rank-2 tensor.

It can be noticed that these expressions can be greatly simplified
if we disregard the contribution of the drift velocity
$\mathbf{v}_{e}$ in the distribution $f_{\mathbf{k}}(t)$ (the
kinetic drift energy is smaller than the thermal energy for any
intensity of the electric field [43,44], and then, because of the
spherical symmetry in the expressions for the tensorial kinetic
coefficients, they become scalars.

Moreover, in Eqs. (70), (71) and (72) are present generalizations of
the so-called Maxwell time [27,28], $\theta_{I_{n_{e}}}$,
$\theta_{h_{e}}$, $\theta_{I_{h_{e}}}$, given by
%
\begin{equation}
\lbrack \theta_{\mathbf{I}_{n_{e}}}(t)]^{-1} = \frac{\beta \hbar
}{m^{\ast }} \sum \limits_{\mathbf{k},\mathbf{q}}\mathbf{q} \cdot
\mathbf{k}A_{\mathbf{kq}} f_{\mathbf{k}} (1 - f_{\mathbf{k}}) \, ,
\end{equation}
%
\begin{equation}
\lbrack \theta_{h_{e}}(t)]^{-1} = \sum
\limits_{\mathbf{k},\mathbf{q}} A_{\mathbf{kq}} b_{2e}
(\epsilon_{\mathbf{k}+\mathbf{q}} - \epsilon_{\mathbf{k}}) \, ,
\end{equation}
%
\begin{equation}
\lbrack \theta_{I_{h_{e}}}(t)]^{-1} = - \sum
\limits_{\mathbf{k},\mathbf{q}} A_{\mathbf{kq}} \mathbf{b}_{4e}
(\epsilon_{\mathbf{k}} \nabla_{\mathbf{k}} \epsilon_{\mathbf{k}} -
\epsilon_{\mathbf{k}+\mathbf{q}} \nabla_{\mathbf{k}+ \mathbf{q}}
\epsilon_{\mathbf{k}+\mathbf{q}}) \, ,
\end{equation}
where
%
\begin{eqnarray}
A_{\mathbf{kq}} &=& \frac{2\pi}{\hbar}
|\mathcal{C}_{\mathbf{kq}}|^{2} \{ [ (\nu_{\mathbf{q}} + 1)
f_{\mathbf{k}+\mathbf{q}} + \nu_{\mathbf{q}} (1 -
f_{\mathbf{k}+\mathbf{q}})] \delta (\epsilon_{\mathbf{k}+\mathbf{q}} -
\epsilon_{\mathbf{k}} - \hbar \omega_{\mathbf{q}}) +  \notag \\
&& [ (\nu_{\mathbf{q}} + 1) (1 - f_{\mathbf{k}+\mathbf{q}}) +
\nu_{\mathbf{q}} f_{\mathbf{k}+\mathbf{q}}] \delta
(\epsilon_{\mathbf{k}+\mathbf{q}} - \epsilon_{\mathbf{k}} + \hbar
\omega_{\mathbf{q}}) \, ,
\end{eqnarray}
$b_{2e}$ is given in Eq. (78) and
%
\begin{equation}
\mathbf{b}_{4e} = [ \Delta_{34}]^{-1} [A_{33} \epsilon_{\mathbf{k}}
\nabla_{\mathbf{k}} \epsilon_{\mathbf{k}} - A_{34}
\nabla_{\mathbf{k}} \epsilon_{\mathbf{k}}] f_{\mathbf{k}} (1 -
f_{\mathbf{k}}) \, ,
\end{equation}
%
\begin{equation}
\Delta _{34} = A_{33} A_{44} - A_{34} A_{43} \, ,
\end{equation}
%
\begin{equation}
A_{33} = \sum_{\mathbf{k}} |\nabla _{\mathbf{k}}
\epsilon_{\mathbf{k}}|^{2} f_{\mathbf{k}} (1 - f_{\mathbf{k}}) \, ,
\end{equation}
%
\begin{equation}
A_{44} = \sum_{\mathbf{k}}(\epsilon_{\mathbf{k}})^{2}
|\nabla_{\mathbf{k}} \epsilon_{\mathbf{k}}|^{2} f_{\mathbf{k}}(1 -
f_{\mathbf{k}}) \, ,
\end{equation}
%
\begin{equation}
A_{34} = \sum_{\mathbf{k}} \epsilon_{\mathbf{k}}| \nabla
_{\mathbf{k}} \epsilon_{\mathbf{k}}|^{2} f_{\mathbf{k}}(1 -
f_{\mathbf{k}}) = A_{43} \, .
\end{equation}

Neglecting the dependence on time of all the different coefficients
(i.e., taken them as weakly dependent on time), going over direct
space, the basic equations of the MHT of order 1 of the carriers in
doped semiconductors are
%
\begin{equation}
\frac{\partial }{\partial t}n_{e}(\mathbf{r},t) + \nabla \cdot
\mathbf{I}_{n_{e}}(\mathbf{r},t) = \Phi _{e}(\mathbf{r},t) \, ,
\end{equation}
%
\begin{eqnarray}
\frac{\partial}{\partial t}\mathbf{I}_{n_{e}}(\mathbf{r},t) &=&
B_{1e}^{[2]}(t)\nabla n_{e}(\mathbf{r},t) - \theta
_{I_{n_{e}}}^{-1}\mathbf{I}_{n_{e}}(\mathbf{r},t) +
b_{34}(t)\mathbf{I}_{h_{e}}\,(\mathbf{r},t) + \notag
\\
&& + b_{34e}^{[2]}(t)\mathbf{I}_{h_{e}}(\mathbf{r},t) +
\frac{e\mathbf{E}}{m_{e}^{\ast}}n_{e}(\mathbf{r},t) \, ,
\end{eqnarray}
%
\begin{eqnarray}
\frac{\partial }{\partial t}h_{e}(\mathbf{r},t) + \nabla \cdot
\mathbf{I}_{h_{e}}(\mathbf{r},t) &=& - \theta
_{h_{e}}^{-1}h_{e}(\mathbf{r},t) + b_{21e}n_{e}(\mathbf{r},t) +  \notag \\
&& + e\mathbf{E} \cdot
\mathbf{I}_{n_{e}}(\mathbf{r},t)n_{e}(\mathbf{r},t) \, ,
\end{eqnarray}
%
\begin{eqnarray}
\frac{\partial }{\partial t}\mathbf{I}_{h_{e}}(\mathbf{r},t) &=& -
C_{2e}^{[2]}\nabla h_{e}(\mathbf{r},t) - \theta
_{I_{h_{e}}}^{-1}\mathbf{I}_{h_{e}}(\mathbf{r},t) -
C_{1e}^{[2]}\nabla n_{e}(\mathbf{r},t) +  \notag \\
&& + b_{43e}(t)\mathbf{I}_{n_{e}}(\mathbf{r},t) +
B_{1e}^{[2]}\mathbf{E}n_{e}(\mathbf{r},t) +
B_{2e}^{[2]}\mathbf{E}h_{e}(\mathbf{r},t) \, .
\end{eqnarray}

In Eq. (92), on the left-hand side is present the barycentric time
differentiation (the conservation part), and on the right the source
of local variations due to the presence of impurities, boundaries,
etc. (we recall that the integration in space of it is null because
the conservation in the number of charges). Equation (93) is on the
right composed of a first term of a diffusive character, followed by
Maxwell contribution, the third contribution is a cross term
associated to thermo-striction effects, and the last one accounts
for the effect of the presence of the electric field creating the
electric current.

In Eq. (94), the left hand side represents the conserving part of
the energy, and on the right we first find Maxwell contribution
which is followed by a cross term associated to thermo-electric
effects, and a contribution due to the presence of the electric
field. The last one is the local production of Joule heat.

In Eq. (95) various terms contribute on the right: the first is of a
diffusive character, followed by Maxwell contribution. The third and
fourth terms are cross terms associated to thermo-striction effects,
and the last two are contribution due to the presence of the
electric field.

We consider next the associated hydrodynamic modes.

\section{The Hydrodynamic Modes in MHT [1]}

For the purpose of obtaining the hydrodynamic modes of the carriers
in the MHT of order 1, we consider Eqs. (69) to (72), but
introducing the simplifications of neglecting the source $\Phi_{e}$
in Eq. (69), i.e., disregarding the effect of impurities and
imperfections. In the last term in Eq. (71) we take for
$n_{e}(\mathbf{r},t)$ only the relevant constant uniform
contribution $n^{0}$, i.e., the doping concentration, and we take
the second-rank tensors, $B_{1e}^{[2]}$, $B_{2e}^{[2]}$,
$C_{1e}^{[2]}$ and $C_{2e}^{[2]}$, as scalars, all of this to have
manageable equations for just to better visualize the physical
characteristics of the hydrodynamic motion.

Transforming Fourier in time Eqs. (69) to (72), we are left with a
set of linear algebraic equations (in $\mathbf{Q}$-$\omega$ space)
whose secular determinant is
%
\begin{equation}
\begin{vmatrix}
i \omega          & -i \mathbf{Q}    &    0         & 0 \\
-B_{1}i\mathbf{Q}\frac{e\mathbf{E}}{m_{e}^{\ast}} & i\omega +\theta
_{I_{n}}^{-1} & -B_{2}i\mathbf{Q} & -b_{34} \\
-b_{21} & -en^{0}\mathbf{E} & i\omega +\theta _{h}^{-1} & -i\mathbf{Q} \\
-C_{1}i\mathbf{Q}-B_{1}\mathbf{E} & -b_{43} &
-C_{2}i\mathbf{Q}-B_{2}\mathbf{E} & i\omega + \theta _{I_{h}}^{-1}
\end{vmatrix}
\end{equation}

The complete set of hydrodynamic modes are the solutions, say
$\omega_{1,2,3,4}$, of a fourth-order algebraic equation, which
follows after making this determinant equal to zero, which we omit
to write down explicitly. We consider now a situation when
thermo-electric effects can be neglected, and then the movements of
density and energy are decoupled.

We do have for the modes associated to the density,
$n(\mathbf{r},t)$, the characteristic equation
%
\begin{equation}
i\omega (i\omega + \theta_{I_{n}}^{-1}) - i\mathbf{Q} \cdot \left(
B_{1}i\mathbf{Q} + \frac{e\mathbf{E}}{m_{e}^{\ast }}\right) =0 \, ,
\end{equation}
or
%
\begin{equation}
\omega^{2} -i \omega \theta_{I_{n}}^{-1} - B_{1}Q^{2} = 0 \, ,
\end{equation}
after neglecting the term with $i\mathbf{Q} \cdot \mathbf{E}$ (we
recall that the electric field is constant and then its divergence
is null). Solution of Eq. (98) provides us with the two roots,
%
\begin{equation}
\omega_{\pm} = \frac{i}{2} \theta_{I_{n}}^{-1} \pm \frac{1}{2}
\sqrt{4B_{1}Q^{2} - \theta_{I_{n}}^{-2}} \, ,
\end{equation}
which can be rewritten as
%
\begin{equation}
\omega_{\pm}(Q) = \frac{i}{2} \theta_{I_{n}}^{-1} \pm \frac{1}{2}
\theta_{I_{n}}^{-1} \sqrt{4B_{1} \theta_{I_{n}}^{2}Q^{2}-1} \, ,
\end{equation}

From this Eq. (100) we can characterize two types of movement:

1. $4B_{1}Q^{2} \theta_{I_{n}}^{2} < 1$, an overdamped regime, when
$\omega_{\pm}$ are purely imaginary, and in the limit $4B_{1}Q^{2}
\theta_{I_{n}}^{2}\ll 1$, we can write
%
\begin{equation}
\omega_{\pm}(Q) = - \frac{i}{2} \theta_{I_{ne}}^{-1} \pm \frac{1}{2}
\theta_{I_{ne}}^{-1} \left( 1-2B_{1} \theta_{I_{n}}^{2}Q^{2} \right)
\, ,
\end{equation}
and then
%
\begin{equation}
\omega_{+}(Q) \simeq -i D_{I_{ne}} Q^{2} \, ,
\end{equation}
%
\begin{equation}
\omega_{-}(Q) = -i \theta_{I_{ne}}^{-1} + i D_{I_{ne}}Q^{2} \, ,
\end{equation}
where $D_{I_{ne}} = B_{1} \theta_{I_{n_{e}}}^{2}$ is a diffusion
coefficient. Hence, the hydrodynamic movement for $Q$ sufficiently
small is of the diffusive type.

2. $4B_{1}Q^{2} \theta_{I_{n}}^{2} > 1$; then $\omega_{\pm}(Q)$ have
an oscillating part and a relaxation time $\theta _{I_{ne}}$. For
$4B_{1}Q^{2} \theta_{I_{n}}^{2} \gg 1$, we obtain
%
\begin{equation}
\omega_{\pm}(Q) \simeq - \frac{i}{2} \theta_{I_{ne}}^{-1} \pm
\frac{1}{2} v_{I_{ne}}Q \, ,
\end{equation}
with $v_{I_{ne}} = \sqrt{B_{1}} \theta_{I_{ne}}^{2}$ having
dimension of velocity. Hence, the hydrodynamic movement for
sufficiently large $Q$ is of the type of a damped wave, where $v$ is
the group velocity of the wave, and the dispersion spectrum is
linear in the wavenumber.

It can be noticed that for any fluid a transition from one regime to
the other (diffusion and damped wave) follows at a cut-off $Q_{co}$
given by $Q_{co}^{2} = (4B_{1} \theta_{I_{n}}^{2})^{-1}$. Movements
well characterized by small wavenumbers ($Q < Q_{co}$), are well
described in MHT of order zero (the classical-Onsagerian one), i.e.
by a Fick diffusion equation. Movements characterized by wavenumbers
$Q > Q_{co}$, are well describe in a MHT of order 1, implying in a
damped wave equation (Maxwell-Cattaneo equation). This is up to a
second cut off wavenumber, say $Q_{12}$, requiring for movements
involving $Q > Q_{12}$ to go over a description in MHT of order 2
[38].

On the other hand, for the modes associated to the thermal motion we
do have
%
\begin{equation}
(i\omega + \theta_{h}^{-1})(i\omega + \theta _{I_{h}}^{-1}) +
C_{2}Q^{2} = 0 \, ,
\end{equation}
which can be written as
%
\begin{equation}
\omega^{2} - i\omega \tau_{h}^{-1} - \tilde{\tau}_{h}^{-2} -
C_{2}Q^{2} = 0 \, ,
\end{equation}
where
%
\begin{equation}
\tau_{h}^{-1} = \theta_{h}^{-1} + \theta_{I_{h}}^{-1} \quad ; \quad
\tilde{\tau}_{h}^{-2} = \theta_{h}^{-1} \cdot \theta_{I_{h}}^{-1} \,
.
\end{equation}

The roots of Eq. (106) are
%
\begin{equation}
\omega_{\pm}(Q) = \frac{i}{2} \tau_{h}^{-1} \pm \frac{1}{2}
\sqrt{C_{2}Q^{2} + \tilde{\tau}_{h}^{-2} - \tau_{h}^{-2}} \, ,
\end{equation}
or
%
\begin{equation}
\omega_{\pm}(Q) = \frac{i}{2} \tau_{h}^{-1} \pm
\frac{\tau_{h}^{-1}}{2} \sqrt{\mathcal{A}-1} \, ,
\end{equation}
where
%
\begin{equation}
\mathcal{A} = 4C_{2} \tau _{h}^{2}Q^{2} + 4\tilde{\tau}_{h}^{-2}
\tau _{h}^{2} \, .
\end{equation}

Quite similarly to the case of charge motion we just considered, we
can evidence two regimes, namely
\begin{itemize}
\item for $\mathcal{A} < 1$, a purely diffusive regime,

\item for $\mathcal{A} > 1$, a damped wave regime,
\end{itemize}
with a cut-off wavenumber $Q_{co}$ defining the frontier between
both given by
%
\begin{equation}
Q_{co}^{2} = (1 - 4 \tilde{\tau}_{h}^{-2} \tau_{h}^{2})/4C_{2}
\tau_{h}^{2} \, ,
\end{equation}
for values of $Q < Q_{co}$ there follows diffusive motion, and for
$Q > Q_{co}$ damped wave motion. This is valid for any fluid, and,
for example, can be visually observed in experiments on thermal
stereolithography (or infrared laser induced rapid phototyping)
[45].

\section{Charge Motion: Electric Conductivity}

In Eq. (93), in the steady state and taking as null $b_{34}$ and
$B_{1}^{[2]}$ meaning that we disregard thermo-striction and
diffusion effects, after integration in space we do obtain that
%
\begin{equation}
e \mathbf{I}_{n_{e}} = \sigma \mathbf{E} \, ,
\end{equation}
where
%
\begin{equation}
\sigma = n_{0} e^{2} \theta_{I_{n_{e}}} / m_{e}^{*} \, ,
\end{equation}
is the usual Sommerfeld-Drude expression for the electric
conductivity; $e \mathbf{I}_{n_{e}}$ is the electric current density
and Eq. (112) is Ohm law.

On the other hand, looking for the space dependence of the current,
after differentiating on time Eq. (93) and using Eq. (92) there
follows that
%
\begin{eqnarray}
\frac{\partial^{2}}{\partial t^{2}}\mathbf{I}_{n_{e}}(\mathbf{r},t)
&=& B_{1}^{[2]e}\nabla \left[ \nabla \cdot
\mathbf{I}_{n_{e}}(\mathbf{r},t) \right] - \frac{1}{\theta
_{I_{ne}}}\frac{\partial }{\partial t}\mathbf{I}
_{n_{e}}(\mathbf{r},t) + \notag \\
&& \frac{e\mathbf{E}}{m_{e}^{\ast}}\nabla \mathbf{\cdot
I}_{n_{e}}(\mathbf{r},t)+\mathbf{G}_{e}(\mathbf{r},t) \, ,
\end{eqnarray}
resembling a Maxwell-Cattaneo-like equation and the telegraphist
equation of electrodynamics, and $\mathbf{G}_{e}(\mathbf{r},t)$ is
the contribution arising out of the term $\Phi $ in the evolution
equation for the density, namely
%
\begin{equation}
\mathbf{G}_{e}(\mathbf{r},t) = \left( B_{1}^{[2]e}\nabla
+\frac{e\mathbf{E}}{m_{e}^{\ast}}\right) \Phi (\mathbf{r},t) \, .
\end{equation}

In the steady state and assuming isotropy such that $B_{1}^{[2]e} =
B_{1e}1^{[2]}$ after multiplying by $\theta_{I_{ne}}$, Eq. (114)
becomes
%
\begin{equation}
R_{e}\nabla^{2}\mathbf{I}_{n_{e}}(\mathbf{r}) +
S_{e}\mathbf{E}\nabla \cdot \mathbf{I}_{n_{e}}(\mathbf{r}) +
\theta_{I_{ne}} \mathbf{G}_{e}(\mathbf{r}) = 0 \, ,
\end{equation}
with $R_{e} = \theta_{I_{ne}}B_{1e}$ and $S_{e} = e
\theta_{I_{ne}}/m_{e}^{\ast}$.

The general solution of Eq. (114) is a sum of solutions of the
associated homogeneous one (obtained for $\mathbf{G}_{e}(\mathbf{r})
\equiv 0$) and a particular solution with
$\mathbf{G}_{e}(\mathbf{r}) \neq 0$.

Consider first the solution of the homogeneous equation. Taking $z$
as the direction along the axis of the cylinder, and the electric
field $\mathbf{E}$ parallel to it, and in cylindrical coordinates
neglecting the dependence on the angle $\theta$, and introducing a
separation in variables $r$ and $z$, we have, after writing
\begin{equation*}
I_{z}(r,z) = I_{z}^{(1)}(r) \times I_{z}^{(2)}(z) \, ,
\end{equation*}
that
%
\begin{equation}
\frac{1}{I_{z}^{(1)}(r)}\frac{1}{r}\frac{\partial}{\partial r}\left[
r\frac{
\partial}{\partial r}I_{z}^{(1)}(r)\right] + \frac{1}{I_{z}^{(2)}(z)}\left[
\frac{\partial^{2}}{\partial z^{2}}I_{z}^{(2)}(z) +
\frac{S_{e}\mathcal{E}}{R_{e}}\frac{\partial }{\partial
z}I_{z}^{(2)}(z)\right] = 0 \, ,
\end{equation}
whose solution [46] is a sum in $\gamma$ of terms like
%
\begin{equation}
I_{z}(r,z) = J_{0}(\gamma ,r) \left[ A_{\gamma} e^{k_{+}(\gamma)z} +
B_{\gamma} e^{k_{-}(\gamma)z} \right] \, ,
\end{equation}
where $\gamma$ is a real number to be determined by the use of
boundary conditions, $J_{0}$ is Bessel function of the second kind,
and $k_{\pm}(\gamma)$ are the roots of
%
\begin{equation}
k^{2} + \frac{S_{e}\mathcal{E}}{R_{e}}k + \gamma^{2} = 0 \, ,
\end{equation}
i.e.,
%
\begin{equation}
k_{\pm}(\gamma) = - \frac{S_{e}\mathcal{E}}{2R_{e}}\pm \frac{1}{2}
\sqrt{\left( \frac{S_{e}\mathcal{E}}{R_{e}}\right) ^{2} -
4\gamma^{2}} = 0 \, .
\end{equation}

Notice that for $\gamma^{2} \gg S_{e} \mathcal{E}/R_{e}$, $k_{\pm}
\rightarrow i\gamma$ and the $\gamma$-dependence of the solutions is
oscillatory. For $\gamma^{2}\ll S_{e}\mathcal{E}/R_{e}$ the two
roots become $k_{+}=0$ and $k_{-}=-S_{e}\mathcal{E}/R_{e}$; hence
the $\gamma$-dependence is ``overdamped" for $k_{-}$ and independent
of $\gamma$ for $k_{+}$. For $ z \gg R_{e} / S_{e}\mathcal{E} >
\gamma $ and $\gamma r \ll 1 $  the solution become independent of
$z$ and $r$. Then, making $A_{\gamma} = \sigma \mathcal{E}$ one
recovers Eq. (112).

In general a non-uniform current distribution may follow from the
space-dependent effects present in the term $G$ of Eqs. (114) and
(115). These are, as already noticed, the distribution of
impurities, presence of imperfections, influence of weldings, and in
the case of nanometric dimensions the question of boundary
conditions in the presence of rugous walls with fractal on average
topography, leading to a complicate reflection of the carriers [47].
Then the coefficients $A_{\gamma}$ and $B_{\gamma}$ in Eq. (118) may
have to be adjusted for the complete solution to satisfy given
boundary conditions. These situations are quite difficult to deal
with theoretically, comprising a case of the so called "hidden
constraints" in systems with complex structure [48]. An analysis of
this question, i.e., the presence of $\mathbf{G}$ and complex
boundary conditions, together with a study of the transient regime
shall be reported in a future communication.

Finally, according to the results here presented, a priori it
appears that the conductivity is weakly dependent on the radius of
the cylinder, but it is limited: The hydrodynamic treatment we have
presented involve the motion of the average of a number of particles
in a volume element, say $d^{3}r$, around position $\mathbf{r}$.
Hence, the results can not be extrapolated to systems with very
short nanometer dimensions, that is, involving lengths comprising a
few lattice parameters: considering a, say, 5 \AA\ lattice parameter
it can be suggested that the results are valid only for lengths
larger than 50 to 100 nm. For smaller distances the motion would be
greatly constrained and, as a rule, with the conductivity becoming
much smaller than the one in bulk.

\section{Heat Motion of Carriers and Phonons}

The subject has been dealt with and reported in Refs. [31], and here
we summarize the results for the sake of completeness of the topic.
We consider Eqs. (94) and (95) for the carriers' density of energy
and its first flux, i.e., we take in direct space
%
\begin{equation}
\frac{\partial }{\partial t}h_{e}(\mathbf{r},t) + \nabla \cdot
\mathbf{I}_{h_{e}}(\mathbf{r},t) = n_{0}e\mathbf{E}\cdot
\mathbf{I}_{n_{e}}(\mathbf{r},t) -
\theta_{h}^{-1}h_{e}(\mathbf{r},t) \, ,
\end{equation}
%
\begin{equation}
\frac{\partial}{\partial t} \mathbf{I}_{h_{e}}(\mathbf{r},t) = -
C_{2}^{[2]e} \nabla h_{e}(\mathbf{r},t) -
\theta_{I_{he}}^{-1}\mathbf{I}_{h_{e}}(\mathbf{r},t) + B_{2}^{[2]e}
\mathbf{E}h_{e}(\mathbf{r},t) \, ,
\end{equation}
where we have neglected thermo-electric effects, that is, we have
taken $b_{21}^{e}$, $b_{43}^{e}$, and $B_{1}^{[2]e}$ as null. In the
\emph{steady state} they become
%
\begin{equation}
\nabla \cdot \mathbf{I}_{h_{e}}(\mathbf{r}) = n_{0}e\mathbf{E} \cdot
\mathbf{I}_{n_{e}}(\mathbf{r}) - \theta_{h}^{-1}h_{e}(\mathbf{r}) \,
,
\end{equation}
%
\begin{equation}
C_{2}^{e}\nabla h_{e}(\mathbf{r}) =
B_{2}^{e}\mathbf{E}h_{e}(\mathbf{r}) - \theta
_{I_{h}}^{-1}\mathbf{I}_{h_{e}}(\mathbf{r}) \, ,
\end{equation}
where we have taken the tensors $B_{2}^{e}$ and $C_{2}^{e}$ as
scalars.

From Eq. (122) we do have for the the heat current that
%
\begin{equation}
\mathbf{I}_{h_{e}}(\mathbf{r}) = - C_{2}^{e} \theta_{I_{h}} \nabla
h_{e}(\mathbf{r}) + B_{2}^{e} \theta_{I_{n_{e}}}
\mathbf{E}h_{e}(\mathbf{r}) \, ,
\end{equation}
and taking into account that the density of energy $h_{e}$ can be
written as
%
\begin{equation}
h_{e}(\mathbf{r}) \approxeq
\frac{3}{2}n_{0}k_{B}T_{e}^{\ast}(\mathbf{r}) +
\frac{n_{0}}{2}m_{e}^{\ast}v_{e}^{2}(\mathbf{r}) \, ,
\end{equation}
being composed of the thermal energy characterized by the
nonequilibrium quasitemperature $T_{e}^{\ast}$, and of the kinetic
energy involving the drift-carrier velocity $v_{e}$. The latter as a
general rule is smaller than the thermal energy [49] and
disregarding it we can write
%
\begin{equation}
\mathbf{I}_{h_{e}}(\mathbf{r}) = -\varkappa_{e}\nabla T_{e}^{\ast}
(\mathbf{r}) + L_{h_{e}}\mathbf{E}\,,
\end{equation}
where
%
\begin{equation}
\varkappa_{e} = \frac{3k_{B}}{2} n_{0} \theta_{I_{he}} \, ,
\end{equation}
can be interpreted as the carriers thermal conductivity which is
space independent, and
%
\begin{equation}
L_{h_{e}} = \frac{3k_{B}}{2}n_{0}B_{2} \theta_{I_{n_{e}}}
T_{e}^{\ast} \, ,
\end{equation}
can be considered as the carriers' thermo-electric coefficient.
Moreover
%
\begin{equation}
C_{2}^{e} = - \frac{15 \hbar^{2}}{4m_{e}^{\ast}} [k_{B}
T_{e}^{\ast}]^{2} \, ,
\end{equation}
and
%
\begin{equation}
B_{2}^{e} = \frac{2 \hbar^{2}}{m_{e}^{\ast}} \, .
\end{equation}

On the other hand, the thermal transport by phonons has been
considered elsewhere, however in intrinsic semiconductors [31]. In
doped semiconductors the influence of the electric field on the
distribution of phonons is presented in Ref. [43], where it is shown
the presence of a kind of resonance (overheating of certain reduced
number of phonon modes in an off-center region of the Brillouin
zone), which is not particularly relevant, arising out of the
process of drifting electron excitation [44]. Therefore, we can
state that the phonons' thermal conductivity is very weakly affected
by the presence of the electric field.

\begin{figure}[b]
\center
\includegraphics[width=8.5cm]{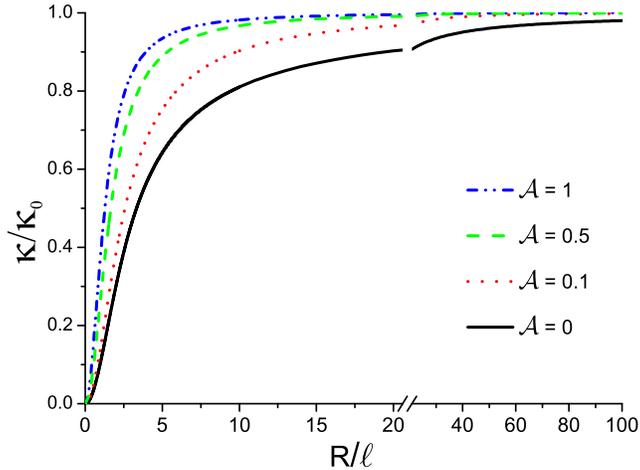}
\caption{Dependence on the scaled wire radius, $R/\ell$, of the
scaled thermal conductivity, $\kappa / \kappa_{0}$, for several
values of the reflection coefficient $\mathcal{A}$ and $\ell^{2} =
s^{2} \theta_{h} \theta_{I_{h}}$, after Ref. [31].}
\end{figure}

The phonons' thermal conductivity is strongly affected by the value
of the radius of the cylinder in the nanometer domain. In Fig. 1 it
is shown such dependence. Parameter $\ell$, with dimensions of
length, is a characteristic length, with $\ell^{2}=s^{2}\theta_{h}
\theta_{I_{h}}$, that is, in a Debye model, its square is given by
the square of the sound velocity times the product of Maxwell times
associated to the phonons energy density and energy flux. Figure 1
tells us that there follows a drastic reduction in thermal
conductivity for $R/\ell$ below the value 10, and becoming orders of
magnitude smaller for $R/\ell <1$. We may then state that in the
range of values of $R/\ell$ there exists a threshold below which the
sample size (radius of the cylinder in units of $\ell$) leads to a
notable reduction of the thermal conductivity, and large increase of
the figure-of-merit in thermo-electric engineering. Figure 1
provides information on the influence of the reflection effect at
the side boundaries: as expected with increasing reflection
coefficient $\mathcal{A}$ there follows an increase in thermal
conductivity. It must be noticed that we have considered normal
reflection at an smooth surface, but the surface is always rugous
with characteristics fractal on average [39] what affects the
reflection processes.

\begin{figure}[b]
\center
\includegraphics[width=8.5cm]{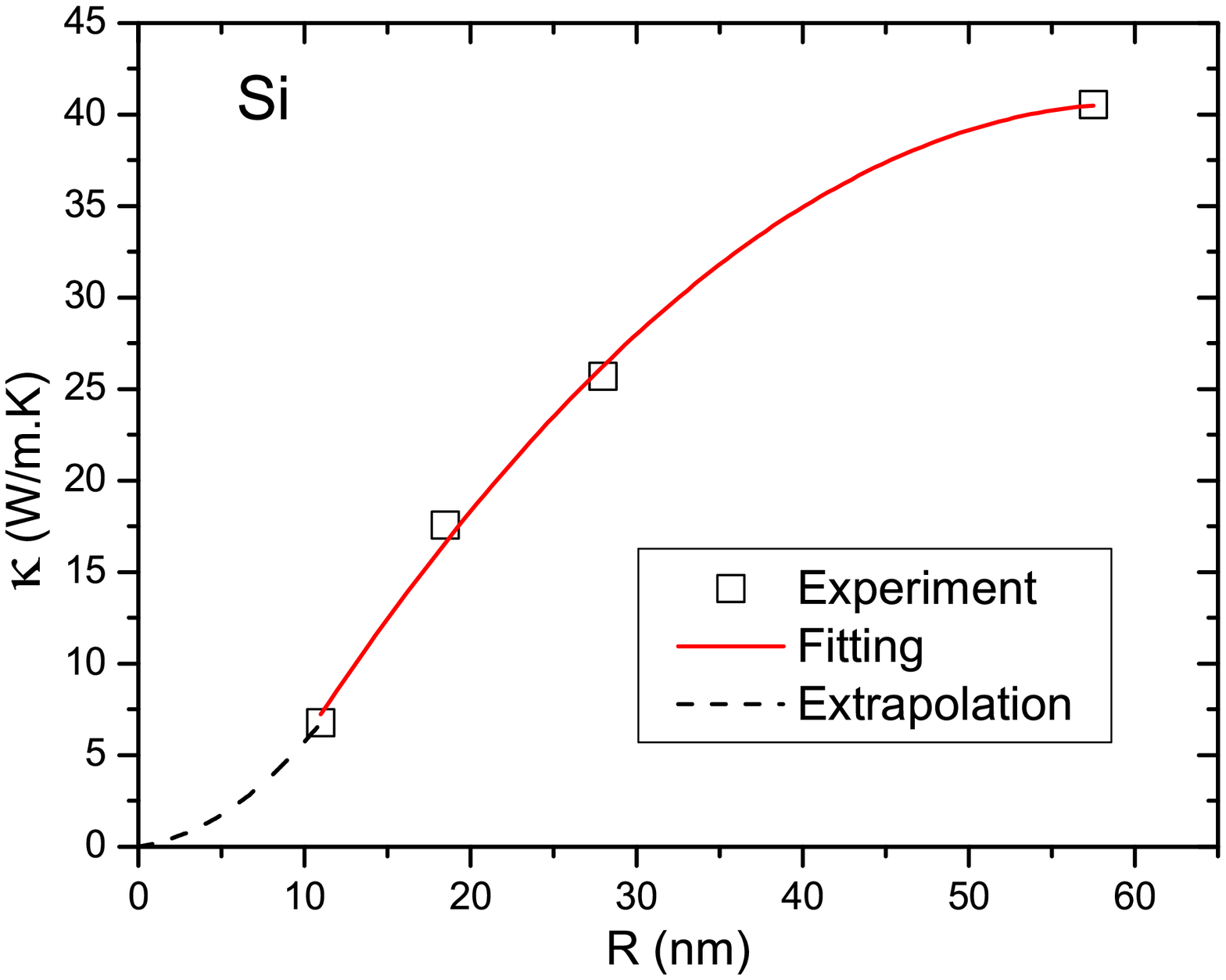}
\caption{Measured thermal conductivity $\kappa$ of wires of Si in
terms of the radius of the wire $R$, at 300 Kelvin; experimental
results ($\Box$) from Ref. [50]; after Ref. [31].}
\end{figure}

Taking into account the experimental data reported by D. Li et al.
(Fig. 1(a) in Ref. [50], where it is shown the measured thermal
conductivity of silicon in terms of the temperature) in samples of
Si nanowires with different diameters (diameters of 22; 37; 56; and
115 nm), we consider those at 300 Kelvin, what is shown in Fig. 2.
If we admit that for all the four samples $\kappa_{0}$ is
approximately the same and of the order of the thermal conductivity
in bulk, namely $\kappa_{0} \simeq$ 148 (W/K.m) [51], we can obtain
the values of $\kappa / \kappa_{0}$ given in Table I (third column),
and from Fig. 1 (for $\mathcal{A}=0$, i.e., no reflection at the
lateral borders: Couette-like flow) we can evaluate that, roughly,
the corresponding values of $R/\ell$ are those given in the fourth
column, and from them we can estimate the values of $\ell$ shown in
the fifth column. Considering as similar the Maxwell times for
energy and its flux, which are equal in a Debye model, that is,
$\theta_{h} = \theta_{I} = \theta$, we get that $\sqrt{3} \ell/s
\approx \theta$, and taking an average sound velocity of 8433 m/s,
we obtain the values for the Maxwell time displayed in column 6 of
Table I. The experimental data (open square dots) in Fig. 2 are
contained in the curve (full line) adjusted by the second order
polynomial
\begin{equation*}
\kappa \simeq -0.014R^{2} + 1.65R - 9.31 \, ,
\end{equation*}
for $R >$ 10 nm. The traced line for $R <$ 10 nm is an intuitive
extrapolated indication, given by: $\kappa \simeq 0.045R^{2} +
0.122R$.

%
\begin{table}[htbp]
\caption{Results for Si}
\begin{tabular}{cccccc}
\hline\hline $R$ (nm) & $\kappa$ (W/K.m) & \, $\kappa/\kappa_{0}$ \, & \, $R/\ell$ \, & \, $\ell$ (nm) & $\theta$ (ps) \\
\tableline 11.0       & 6.76             & 0.046                     & 0.626          & 17.57          & 3.61 \\
18.5                  & 17.57            & 0.119                     & 1.063          & 17.40          & 3.57 \\
28.0                  & 25.68            & 0.173                     & 1.339          & 20.91          & 4.29 \\
57.5                  & 40.54            & 0.274                     & 1.850          & 31.08          & 6.38 \\
\hline\hline
\end{tabular}
\end{table}
%

\begin{figure}[b]
\center
\includegraphics[width=8.5cm]{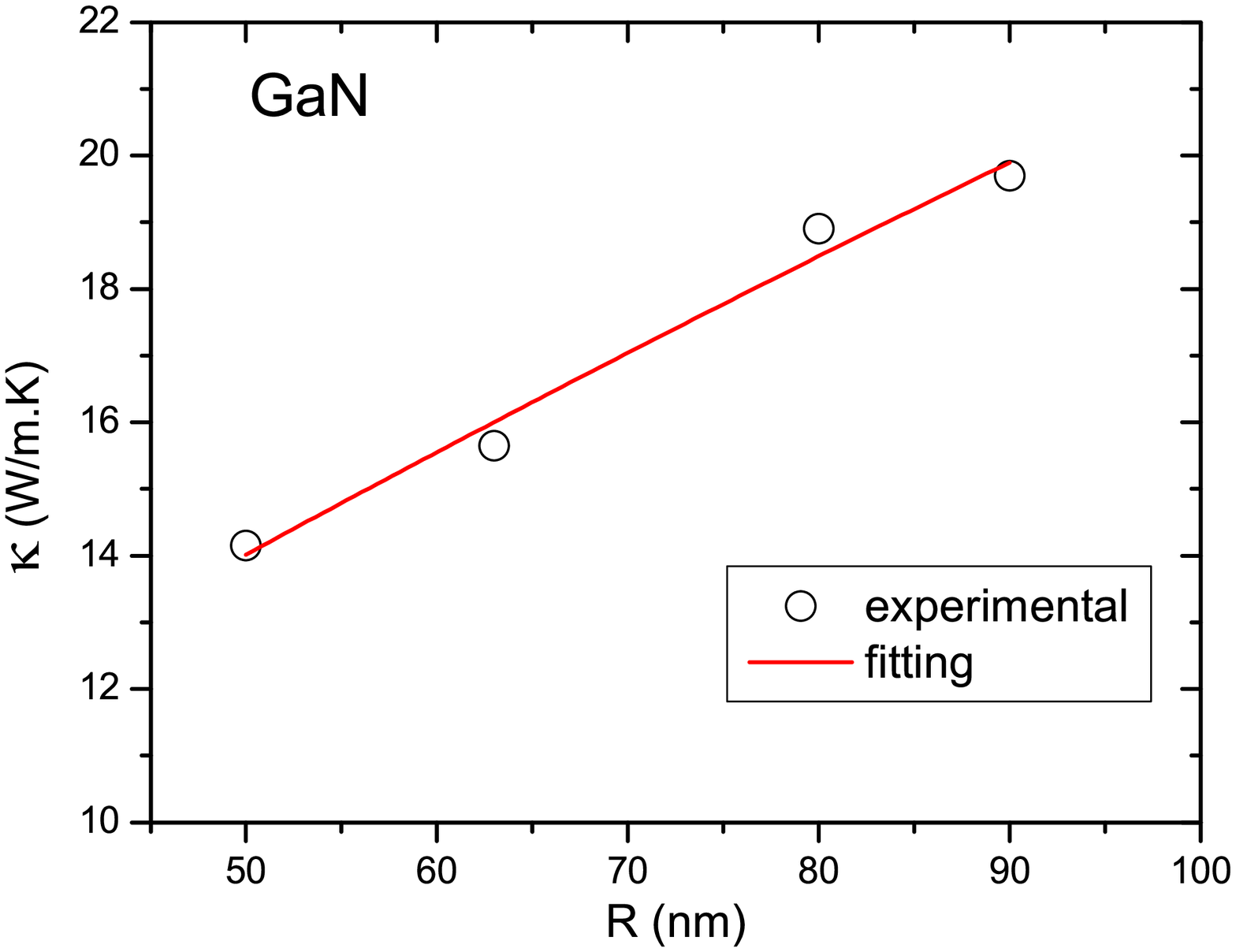}
\caption{Measured thermal conductivity $\kappa$ of wires of GaN in
terms of the radius of the wire $R$, at 300 Kelvin; experimental
results ($\bigcirc$) from Ref. [52]; after Ref. [32].}
\end{figure}

From the experimental data reported by C. Guthy et al. (Fig. 2(a) in
Ref. [52], where it is shown the measured thermal conductivity of
GaN in terms of the temperature) in GaN nanowires with different
diameters (diameters of 100; 126; 160; and 181 nm), we consider
those at 300 Kelvin, what is shown in Fig. 3. Using the value of the
thermal conductivity in bulk for GaN, namely $\kappa_{0} \simeq$ 210
(W/K.m) [53,54] and taking an average (in this hexagonal crystal)
sound velocity of 5170 m/s [52], we obtain, similarly to Table I,
the values shown in Table II. The experimental values (open circular
dots) in Fig. 3 are contained in the curve (full line) adjusted by
the second order polynomial
\begin{equation*}
\kappa \simeq -0.0002R^{2} + 0.18R + 5.7 \, ,
\end{equation*}
for $R >$ 50 nm.

%
\begin{table}[htbp]
\caption{Results for GaN}
\begin{tabular}{cccccc}
\hline\hline $R$ (nm) & $\kappa$ (W/K.m) & \, $\kappa/\kappa_{0}$ \, & \, $R/\ell$ \, & \, $\ell$ (nm) & $\theta$ (ps) \\
\tableline 50.0       & 14.1             & 0.067                     & 0.769          & 65.0           & 21.7 \\
63.0                  & 15.6             & 0.074                     & 0.813          & 77.5           & 26.0 \\
80.0                  & 18.9             & 0.090                     & 0.905          & 83.4           & 27.9 \\
90.5                  & 19.7             & 0.093                     & 0.922          & 98.1           & 32.9 \\
\hline\hline
\end{tabular}
\end{table}

It must also be noticed the important point that the characteristic
length $\ell$ and Maxwell times depend on $R$ and on the
nonequilibrium thermodynamic state of the system. This is so because
of their dependence on $R$, which determines the frequencies
$\omega_{nq_{z}}$ and of the sum over $nq_{z}$. Figure 4 shows the
dependence on the wire radius $R$ of the characteristic length
$\ell$ and Maxwell time $\theta$ for GaN and Si nanowires. The
linear expressions that relate the characteristic length and Maxwell
time with the radius $R$ are indicated within the figures.

\begin{figure}[h]
\center
\includegraphics[width=8.0cm]{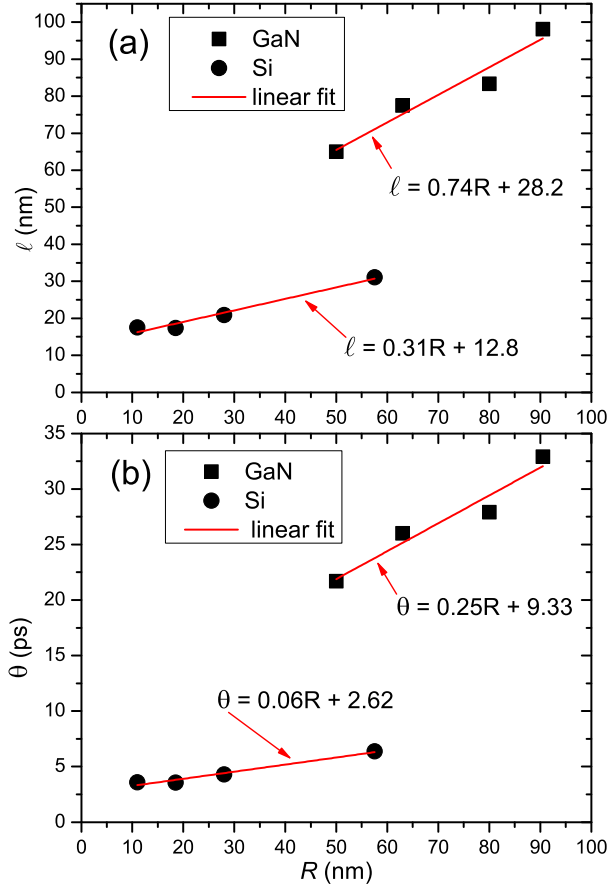}
\caption{Dependence on the wire radius $R$ of the (a) characteristic
length $\ell$ and (b) Maxwell time $\theta$, for GaN and Si
nanowires, after Ref. [31].}
\end{figure}

Concerning the so-called figure of merit, $Z$, which is a number
that allows for obtaining a useful insight for optimizing design
parameters, is constructed by choosing the parameters that are most
centrally vital to a design solution. For the case of
thermo-electric devices is used [3]
%
\begin{equation}
Z = S_{e} \sigma /\kappa \, ,
\end{equation}
where $S_{e}$ is Seebeck coefficient. If we consider Seebeck effect
and the electric conductivity as nearly independent on size, and the
phonon thermal conductivity as the relevant one, we can see that,
according to the results in figures 1 to 3, the figure of merit $Z$
of Eq. (132) greatly increases in quantum wires with radius in the
interval of 10 to 90 nm.

\section{Concluding Remarks}

We have presented an extended theory of the Mesoscopic
Hydro-Thermodynamics of phonons and carriers in n-doped direct gap
polar semiconductors in the presence of electric fields. MHT, also
referred to as Higher-Order Generalized Hydrodynamics, extends
standard (or Onsagerian) hydrodynamics allowing to incorporate
hydrodynamic motion not restricted to smooth in space and time
characteristics (i.e., including intermediate to short wavelengths
and intermediate to high frequencies). It consists in deriving a set
of coupled hydrodynamic equations for the densities of
quasi-particles (carriers and phonons) and of energy and their
fluxes of all orders. This has been done in Section III.

The matter has been illustrated resorting to a MHT of order 1 for
carriers and phonons, which is a \emph{contracted description} in
terms of their densities, energies, and the vectorial fluxes
(electric current and heat current) of both. Criteria for performing
such contraction are discussed in Ref. [38].

The corresponding four hydrodynamic equations are coupled together,
but if we disregard the cross-contributions associated to
thermo-electric effects, there follows the separate sets of two
equations for the motion of charges and two equations for the motion
of energy. These are the basic Eqs. (69) and (70), and Eqs. (71) and
(72), respectively. It may be noticed that in these equations are
present the quite important \emph{generalizations of Maxwell time}.
We recall that the origin of Maxwell time goes back to the
fundamental article by J.C. Maxwell in 1867, on the dynamical theory
of gases and liquids [27], in the strain rate model there presented
it is considered as representing the time during which the stresses
are damped [28]. Section IV is closed with an interpretation of the
several contributions to the hydrodynamic equations.

In Section V the hydrodynamic modes in this MHT of order 1 are
derived. They allow to characterize the two regimes that are covered
by it, namely, a diffusive motion at low wavenumbers and a damped
wave motion at intermediate wavenumbers. In the first case the
motion is governed by a typical diffusion equation (Fick's and
Fourier's type respectively), and in the second by a
Maxwell-Cattaneo-like equation. A cut-off wavenumber,
$Q_{\mathrm{co}}$ of Eq. (100), defines the frontier between the two
types of regimes.

Charge motion and characterization of the electric conductivity in
the steady state are analyzed in Section VI. It may be notice that
at no too small nanometric sizes the conductivity is nearly constant
and taking a Drude-type expression. Minor space-dependent effects
may result from the presence of the space-dependent distribution of
impurities, imperfections, influence of weldings, and boundary
conditions (which have a ruggedness of a fractal-on-average type).
As noticed at the closing of Section VI, the results can not be
extrapolated to wires with very short nanometer radius, say, below a
few tenths of nanometers.

Heat motion and characterization of the thermal conductivity in the
steady state are analyzed in Section VIII. In the case of the
carriers, as it happens with the electric conductivity, the thermal
conductivity is constant with an expression of the type of standard
kinetic theory, and taking into account the expression for the
electric conductivity there follows a type of Wiederman-Franz law.
In the case of the phonons, quite differently, there follows a
strong space dependence affected by the value of the radius of the
wire. There follows a drastic reduction in the thermal conductivity
as the radius decreases, evidenced within this MHT of order 1, which
is being suppressed if one resorts to standard hydrodynamics. This
may be interpreted that as the radius decreases to the nanometric
scale larger wavenumbers need be included for the proper description
of the movement. A MHT of higher order than 1, would be required for
wires with radius of a few nanometers.

Finally, as a consequence of those results we can draw the attention
to the fact that the so-called figure of merit in the engineering of
thermo-electric devices would greatly increase following the
decrease of the wire's radius. \\

\textbf{Acknowledgments:} The authors would like to acknowledge
partial financial support received from the S\~{a}o Paulo State
Research Agency (FAPESP), Goi\'{a}s State Research Agency (FAPEG),
the Brazilian National Research Council (CNPq), and the Brazilian
Synchroton Source (LNLS) under a scientific collaborations agreement
with Unicamp. \vspace{2cm}

\textbf{In Memoriam:} \emph{With very sad feelings, we regret to
report the passing away of our dear colleague \'{A}urea Rosas
Vasconcellos, a genuine, devoted and extremely competent Teacher and
Researcher with fervent dedication to Theoretical Physics in the
Condensed Matter area, who was a quite important contributor to the
development of the present work}.

\newpage

\appendix


\section{The Nonequilibrium Statistical Operator}

According to NESEF ([9,11-13,25] with a short overview given in Ref.
[32]), the nonequilibrium statistical operator in terms of the basic
nonequilibrium variables in sets (8) and (9) is given by
%
\begin{equation}
\mathcal{R}_{\varepsilon}(t) = \varrho_{\varepsilon }(t)\times
\varrho_{B} \; ,
\end{equation}
where
%
\begin{equation}
\varrho_{\varepsilon}(t) = \exp \Big{\{}\ln \bar{\varrho}(t,0) -
\int \limits_{-\infty}^{t}dt^{\prime }e^{\varepsilon (t^{\prime}-t)}
\frac{d}{dt^{\prime}} \ln \bar{\varrho}(t^{\prime},t^{\prime }-t)
\Big{\}} \, ,
\end{equation}
with $\bar{\varrho}(t,0)$ being the auxiliary statistical operator
(also called ``instantaneous quasi-equilibrium operator") and
%
\begin{eqnarray}
\bar{\varrho}(t^{\prime},t^{\prime }-t) &=& \exp \Big{\{} - \phi(t)
- \sum \limits_{\mathbf{k}} F_{\mathbf{k}}(t^{\prime})
\hat{n}_{\mathbf{k}}(t^{\prime}-t) - \sum
\limits_{\mathbf{k},\mathbf{Q} \neq 0}
F_{\mathbf{kQ}}(t^{\prime}) \hat{n}_{\mathbf{kQ}}(t^{\prime}-t)  \notag \\
&& - \sum \limits_{\mathbf{q},\gamma} \varphi_{\mathbf{q\gamma
}}(t^{\prime}) \hat{\nu}_{\mathbf{q\gamma}}(t^{\prime}-t) - \sum
\limits_{\substack{\mathbf{q},\mathbf{Q} \neq 0  \\ \gamma}}
\varphi_{\mathbf{qQ\gamma}}(t^{\prime})
\hat{\nu}_{\mathbf{qQ\gamma}}(t^{\prime }-t) \Big{\}} \; ,
\end{eqnarray}
where $t^{\prime}$ stands for the dependence on the time of the
nonequilibrium thermodynamic variables $F$'s and the dynamical
microvariables, in Heisenberg representation, depend on ($t^{\prime}
- t$). Moreover $\varrho_{B}$ is the canonical distribution of the
bath of acoustic phonons in equilibrium at temperature $T_{0}$, and
$\phi(t)$ ensuring the normalization plays the role of the logarithm
of a nonequilibrium partition function.

We recall that the second term in the exponent in Eq. (A.2) accounts for
historicity and irreversibility in the nonequilibrium state of the system.
The quantity $\varepsilon $ is a positive infinitesimal that goes to zero
after the trace operation in the calculation of averages has been performed.
We also recall that
%
\begin{equation}
\varrho_{\varepsilon}(t) = \bar{\varrho}(t,0) +
\varrho_{\varepsilon}^{\prime}(t) \; ,
\end{equation}
i.e., it has an additive composition property, with a contribution
of the instantaneous quasi-equilibrium statistical operator plus the
one of $\varrho_{\varepsilon}^{\prime}$ which contains the
historicity and produces irreversible evolution.

\section{NESEF-Kinetic Theory}

The NESEF-based Kinetic Theory of relaxation processes basically
consists into taking the average over the nonequilibrium ensemble of
Heisenberg (or Hamilton at the classical level) equations of motion
of the dynamical operators for the observables, say,
$\hat{A}_{j}(\mathbf{r})$, with $j=1,2,...$, (a function over phase
space in classical mechanics and Hermitian operator in quantum
mechanics) under consideration, i.e.
%
\begin{equation}
\frac{\partial}{\partial t} A_{j}(\mathbf{r},t) =
\frac{\partial}{\partial t} \mathrm{Tr} \{\hat{A}_{j}(\mathbf{r})
\varrho_{\varepsilon}(t) \times \varrho_{B} \} = \mathrm{Tr} \left\{
\frac{1}{i\hbar}[\hat{A}_{j}(\mathbf{r}),\hat{H}]\varrho_{\varepsilon}(t)
\times \varrho_{B} \right\} \, ,
\end{equation}
which is a manifestation of Ehrenfest Theorem. The practical
handling of this NESEF-Kinetic Theory is described in Refs.
[9,11-13] and mainly in [26]. The NESEF is a powerful formalism that
provides an elegant, practical, and physically clear picture for
describing irreversible processes, adequate to deal with a large
class of experimental situations, as for example, in semiconductors
far-from equilibrium, obtaining good agreement in comparisons with
other theoretical and experimental results [55].

Here we briefly notice that the Markovian limit of the kinetic
theory is of particular relevance as a result that, for a large
class of problems, the interactions involved are weak and the use of
this lowest order, second order in the interaction strengths, in the
equations of motion constitutes an excellent approximation of good
practical value. By means of a different approach, E. B. Davies [56]
has shown that in fact the Markovian approach can be validated in
the weak coupling (in the interaction) limit.

Explicitly written, the Markovian equations in the kinetic theory are
%
\begin{equation}
\frac{\partial}{\partial t}A_{j}(\mathbf{r},t) =
J_{j}^{(0)}(\mathbf{r},t) + J_{j}^{(1)}(\mathbf{r},t) +
J_{j}^{(2)}(\mathbf{r},t),
\end{equation}
where, after it is introduced in the Hamiltonian the separation
$\hat{H} = \hat{H}_{0} + \hat{H}^{\prime}$, where $\hat{H}_{0}$
stands for the kinetic energy and $\hat{H}^{\prime}$ contains the
interaction potential energies present in Eq. (B1), we have that
%
\begin{equation}
J_{j}^{(0)}(\mathbf{r},t) = \mathrm{Tr}\left\{ \frac{1}{i\hbar
}[\hat{A}_{j}(\mathbf{r}),\hat{H}_{0}]\bar{\varrho}(t,0)\times
\varrho_{R}\right\} \, ,
\end{equation}
%
\begin{equation}
J_{j}^{(1)}(\mathbf{r},t)=\mathrm{Tr}\left\{ \frac{1}{i\hbar
}[\hat{A}_{j}(\mathbf{r}),\hat{H}^{\prime}]\bar{\varrho}(t,0)\times
\varrho_{R}\right\} \, ,
\end{equation}
and $J_{j}^{(2)}(\mathbf{r},t)=$ $_{I}J_{j}^{(2)}(\mathbf{r},t)$ $+$
$_{II}J_{j}^{(2)}(\mathbf{r},t)$, with
%
\begin{equation}
_{I}J_{j}^{(2)}(\mathbf{r},t) = \frac{1}{(i\hbar
)^{2}}\int\limits_{-\infty}^{t}dt^{\prime} e^{\varepsilon
(t^{\prime}-t)} \mathrm{Tr} \left\{ \big[ \hat{H}^{\prime}(t^{\prime
}-t)_{0},[\hat{H}^{\prime},\hat{A}_{j}(\mathbf{r})] \big]
\bar{\varrho}(t,0) \times \varrho_{R}\right\} \, ,
\end{equation}
%
\begin{equation}
_{II}J_{j}^{(2)}(\mathbf{r},t) = \frac{1}{i\hbar
}\sum\limits_{\mathbf{k}}\int\limits_{-\infty }^{t}dt^{\prime
}e^{\varepsilon (t^{\prime}-t)} \mathrm{Tr}\left\{ [\hat{H}^{\prime
},\Hat{A}_{k}(\mathbf{r})]\bar{\varrho} (t,0)\times \varrho
_{R}\right\} \frac{\delta J_{j}^{(1)}(\mathbf{r},t)}{ \delta
A_{k}(\mathbf{r},t)} \, ,
\end{equation}
where $\bar{\varrho}$ is the auxiliary statistical operator of Eq.
(A.3) and $\varrho_{R}$ the equilibrium statistical distribution of
the thermal bath, and we recall that $J_{j}^{(0)}$ and
$J_{j}^{(1)}$, which in Mori's terminology [57] are called the
precession and force terms, are related to the non-dissipative part
of the motion, while dissipative effects are accounted for in
$J_{j}^{(2)}$ which can be called scattering integrals. Subindex
nought indicates evolution in the interaction representation,
$\delta$ indicates functional differentiation [44].

\section{Summary of Heims-Jaynes Procedure}

Given an statistical operator of the form
%
\begin{equation}
\varrho = \frac{1}{Z} e^{\widehat{A} + \widehat{B}} ,
\end{equation}
where
%
\begin{equation}
Z = \mathrm{Tr} \big\{ e^{\widehat{A} + \widehat{B}} \big\} \, ,
\end{equation}
ensures its normalization, and introducing
%
\begin{equation}
\varrho_{0} = \frac{e^{\widehat{A}}}{\mathrm{Tr} \big\{
e^{\widehat{A}}\big\}} \, ,
\end{equation}
according to Heims-Jaynes, given an any operator $\widehat{\Theta}$
it follows that
%
\begin{equation}
\mathrm{Tr}\big\{\widehat{\Theta }\varrho \big\} = \langle
\widehat{\Theta} \rangle _{0} + \sum \limits_{n=1}^{\infty} \langle
\widehat{Q}_{n}(\widehat{\Theta} - \langle \widehat{\Theta} \rangle
_{0})\rangle \, ,
\end{equation}
where
%
\begin{equation}
\langle \widehat{\Theta} \rangle_{0} = \mathrm{Tr} \big\{
\widehat{\Theta} \varrho_{0} \big\} \, ,
\end{equation}
with
%
\begin{equation}
\widehat{Q}_{n} = \widehat{S}_{n} - \sum \limits_{k=1}^{n-1} \langle
\widehat{Q}_{n} \rangle_{0}\widehat{S}_{n-k} \, ,
\end{equation}
for $n\geq 2$, and $\widehat{Q}_{0}=\widehat{1}$ and
$\widehat{Q}_{1} = \widehat{S}_{1}$,
%
\begin{equation}
\widehat{S}_{n} = \frac{B^{n}}{n!} \, , \quad \widehat{S}_{0} =
\widehat{1} \, .
\end{equation}

Equation (C4) consists of the average value of $\widehat{\Theta}$
with $\varrho _{0}$ (that is, only depending on $A$) plus a
contribution in the form of a series expansion in powers of $B$. In
a first-order approximation we do have that
%
\begin{equation}
\mathrm{Tr} \big\{ \widehat{\Theta} \varrho \big\} \simeq \langle
\widehat{\Theta} \rangle_{0} + \mathrm{Tr}
\big\{\widehat{B}(\widehat{\Theta} - \langle \widehat{\Theta}
\rangle_{0}) \varrho_{0}\big\} \, .
\end{equation}

In Section III we have used that
%
\begin{equation}
\overline{\varrho}(t,0) = \frac{1}{\overline{Z}(t)} e^{\widehat{A} +
\widehat{B}} \; .
\end{equation}
where
%
\begin{eqnarray}
A(t) &=& F_{n_{e}}(t)\widehat{n}_{e} + \mathbf{F}_{n_{e}}(t)\cdot
\widehat{\mathbf{I}}_{n_{e}} +  \notag \\
&& F_{h_{e}}(t)\widehat{h}_{e} + \mathbf{F}_{h_{e}}(t)\cdot
\widehat{\mathbf{I}}_{h_{e}} +  \notag \\
&& F_{n_{p}}(t)\widehat{n}_{p} + \mathbf{F}_{n_{p}}(t)\cdot
\widehat{\mathbf{I}}_{n_{p}} +  \notag \\
&& F_{h_{p}}(t)\widehat{h}_{p} + \mathbf{F}_{h_{p}}(t) \cdot
\widehat{\mathbf{I}}_{h_{p}} \; ,
\end{eqnarray}
that is, the homogenous part, $\mathbf{Q}\neq 0$, in the exponent of
Eq. (48), and $B(t)$ is the inhomogeneous part, meaning the
contributions with $\mathbf{Q} \neq 0$, and we have used the
first-order (linear in $\widehat{B}$) approximation.

In particular we had that
%
\begin{equation}
I_{n_{e}}^{[2]}(\mathbf{Q},t) = B_{1e}^{[2]}(t)n_{e}(\mathbf{Q},t) +
B_{2e}^{[2]}(t)h_{e}(\mathbf{Q},t) \, ,
\end{equation}
%
\begin{equation}
I_{h_{e}}^{[2]}(\mathbf{Q},t) = C_{1e}^{[2]}(t)n_{e}(\mathbf{Q},t) +
C_{2e}^{[2]}(t)h_{e}(\mathbf{Q},t) \, ,
\end{equation}
with tensor $B$ and $C$ given in Eqs. (73) to (76).

On the other hand, for the case of the phonons we do obtain that
%
\begin{equation}
\nu _{\mathbf{q}}(t)\simeq \overline{\nu }_{\mathbf{q}}(t) -
\overline{\nu}_{\mathbf{q}}(t)[1 +
\overline{\nu}_{\mathbf{q}}(t)][\mathbf{F}_{n}(t) \cdot
\nabla_{\mathbf{q}} \omega_{\mathbf{q}} + \mathbf{F}_{h}(t) \cdot
\hbar \omega _{\mathbf{q}} \nabla_{\mathbf{q}} \omega_{\mathbf{q}}]
\, , \label{eqc10}
\end{equation}
where
%
\begin{equation}
\overline{\nu }_{\mathbf{q}}(t) = \frac{1}{e^{\{\varphi_{n}(t) +
\varphi_{h}(t) \hbar \omega_{\mathbf{q}}\}}-1} \, , \label{eqc11}
\end{equation}
that is, a first order Taylor expansion in $\mathbf{F}_{n}$ and
$\mathbf{F}_{h}$ (linear approximation).

Next, resorting to the use of the nonequilibrium equations of state
that relate the four nonequilibrium thermodynamic variables to the
four basic variables, it follows in first-order Heims-Jaynes
expansion that
%
\begin{equation}
n(\mathbf{Q},t) = \overline{A} _{11}(t) \varphi_{n}(\mathbf{Q},t) +
\overline{A}_{12}(t) \varphi_{h}(\mathbf{Q},t) \, ,  \label{eqc12}
\end{equation}
%
\begin{equation}
\mathbf{I}_{n}(\mathbf{Q},t) = \overline{A}_{33}^{[2]}(t) \cdot
\mathbf{F} _{n}(\mathbf{Q},t) + \overline{A}_{34}^{[2]}(t) \cdot
\mathbf{F}_{h}(\mathbf{Q},t) \, ,  \label{eqc13}
\end{equation}
%
\begin{equation}
h(\mathbf{Q},t) = \overline{A}_{21}(t) \varphi _{n}(\mathbf{Q},t) +
\overline{A}_{22}(t) \varphi_{h}(\mathbf{Q},t) \, ,  \label{eqc14}
\end{equation}
%
\begin{equation}
\mathbf{I}_{h}(\mathbf{Q},t) = \overline{A}_{43}^{[2] }(t) \cdot
\mathbf{F}_{n}(\mathbf{Q},t) + \overline{A}_{44}^{[2]}(t) \cdot
\mathbf{F}_{h}(\mathbf{Q},t) \, ,  \label{eqc15}
\end{equation}
where $\overline{A}_{11}$, $\overline{A}_{12}$,
$\overline{A}_{33}^{[2]}$, $\overline{A}_{34}^{[2]}$,
$\overline{A}_{21}$, $\overline{A}_{22}$, $\overline{A}_{34}^{[2]}$
and $\overline{A}_{44}^{[2]}$ are those of Eqs. (C24) to (C29)
below, except for the replacement of $\nu_{\mathbf{q}}(t)$ of Eq.
(C13) by $\overline{\nu}_{\mathbf{q}}(t)$ of Eq. (C14).

In Eqs. (C15) and (C17) the contributions in $\mathbf{F}_{n}$ and
$\mathbf{F}_{h}$ present in Eq. (C13) are null, whereas in Eqs.
(C.16) and (C.18) are null the contributions in $\varphi_{n}$ and
$\varphi_{h}$. Eqs. (C15) to (C18) constitute a set of linear
algebraic equations that can be inverted to obtain the four
nonequilibrium thermodynamic variables $\varphi_{n}$, $\varphi_{h}$,
$\mathbf{F}_{n}$ and $\mathbf{F}_{h}$, in terms of the basic
hydrodynamic quantities, $n$, $h$, $\mathbf{I}_{n}$ and
$\mathbf{I}_{h}$.

The second-order fluxes are given by
%
\begin{equation}
I_{n}^{[2]}(\mathbf{Q},t) = \overline{A}_{33}^{[2]}(t)
\varphi_{n}(\mathbf{Q},t) + \overline{A}_{34}^{[2]}(t)
\varphi_{h}(\mathbf{Q},t) \, , \label{eqc16}
\end{equation}
%
\begin{equation}
I_{h}^{[2]}(\mathbf{Q},t) = \overline{A}_{34}^{[2]}(t)
\varphi_{n}(\mathbf{Q},t) + \overline{A}_{44}^{[2]}(t)
\varphi_{h}(\mathbf{Q},t) \, , \label{eqc17}
\end{equation}
where $\overline{A}_{33}^{[2]}$, $\overline{A}_{34}^{[2]}$ and
$\overline{A}_{44}^{[2]}$ are those of Eqs. (C27), (C28) and (C29),
except for the replacement of $\nu_{\mathbf{q}}(t)$ of Eq. (C13) by
$\overline{\nu}_{\mathbf{q}}(t)$ of Eq. (C14).

On the other hand, introducing the concept of nonequilibrium
temperature, better called quasitemperature $T^{\ast
}(\mathbf{r},t)$ in the form
%
\begin{equation}
k_{B}T^{\ast }(\mathbf{r},t) = \frac{1}{\varphi
_{h}(\mathbf{r},t)}\,, \label{eqc18}
\end{equation}
we can obtain an evolution equation for it starting with the
evolution equation for the energy in the form of the hyperbolic
Maxwell-Cattaneo equation, from which together with the
nonequilibrium thermodynamic equation of state, Eq. (C17), we have
that
%
\begin{equation*}
\left[ \sum_{\mathbf{q}}(\hbar \omega
_{\mathbf{q}})^{2}\overline{\nu }_{\mathbf{q}}(t)[1+\overline{\nu
}_{\mathbf{q}}(t)]\right] \left[ \frac{\partial ^{2}\varphi
_{h}(\mathbf{r},t)}{\partial t^{2}} + \left( \theta_{h}^{-1} +
\theta _{\mathbf{I}_{h}}^{-1}\right) \frac{\partial \varphi
_{h}(\mathbf{r},t)}{\partial t} \right] +
\frac{h(\mathbf{r},t)}{\theta_{h}\theta_{\mathbf{I}_{h}}}
\end{equation*}
\begin{equation*}
= \nabla \cdot \left[
\overline{A}_{43}^{[2]}(t)\frac{\overline{A}_{12}(t)}{\overline{A}_{11}(t)}-\overline{A}_{44}^{[2]}(t)
\right] \cdot \nabla \varphi _{h}(\mathbf{r},t) +
\end{equation*}
\begin{equation}
\nabla \cdot \left[
\frac{\overline{A}_{34}^{[2]}(t)}{\overline{A}_{11}(t)} \right]
\cdot \nabla n(\mathbf{r},t) + \theta
_{\mathbf{I}_{h}}^{-1}\mathcal{I}_{h}^{[0]\mathrm{ext.}}(\mathbf{r},t)
\, , \label{eqc19}
\end{equation}
and, after introducing the heat capacity
%
\begin{equation}
C_{V}(t) = k_{B}\sum_{\mathbf{q}}\left( \frac{\hbar \omega
_{\mathbf{q}}}{k_{B}T_{0}}\right)^{2}\overline{\nu}_{\mathbf{q}}(t)
[1 + \overline{\nu}_{\mathbf{q}}(t)] \, , \label{eqc20}
\end{equation}
where $T_{0}$ is the temperature in equilibrium in this linear
treatment, and the quantities $A_{ij}$ and $\Delta _{ij}$ are,
%
\begin{equation}
A_{11}(t) = \sum_{\mathbf{q}} \nu_{\mathbf{q}}(t)[1 +
\nu_{\mathbf{q}}(t)] \, , \label{eqd5}
\end{equation}
%
\begin{equation}
A_{12}(t) = A_{21}(t) = \sum_{\mathbf{q}} \nu_{\mathbf{q}}(t)[1 +
\nu_{\mathbf{q}}(t)] \hbar \omega_{\mathbf{q}} \, ,  \label{eqd6}
\end{equation}
%
\begin{equation}
A_{22}(t) = \sum_{\mathbf{q}} \nu_{\mathbf{q}}(t)[1 + \nu
_{\mathbf{q}}(t)] (\hbar \omega_{\mathbf{q}})^{2} \, ,  \label{eqd7}
\end{equation}
%
\begin{equation}
A_{33}^{[2]}(t) = \sum_{\mathbf{q}} \nu_{\mathbf{q}}(t) [1 +
\nu_{\mathbf{q}}(t)] [\nabla_{\mathbf{q}} \omega
\mathbf{_{\mathbf{q}}} \nabla_{\mathbf{q}} \omega_{\mathbf{q}}] \, ,
\label{eqd8}
\end{equation}
%
\begin{equation}
A_{34}^{[2]}(t) = A_{43}^{[2]}(t) = \sum_{\mathbf{q}} \nu
_{\mathbf{q}}(t)[1 + \nu_{\mathbf{q}}(t)] [\nabla_{\mathbf{q}}
\omega_{\mathbf{q}} \nabla_{\mathbf{q}} \omega_{\mathbf{q}}] \hbar
\omega_{\mathbf{q}} \, ,  \label{eqd9}
\end{equation}
%
\begin{equation}
A_{44}^{[2]}(t) = \sum_{\mathbf{q}} \nu_{\mathbf{q}}(t) [1 +
\nu_{\mathbf{q}}(t)] [\nabla_{\mathbf{q}} \omega_{\mathbf{q}}
\nabla_{\mathbf{q}}\omega_{\mathbf{q}}](\hbar
\omega_{\mathbf{q}})^{2} \, ,  \label{eqd10}
\end{equation}
%
\begin{equation}
\Delta_{12}(t) = A_{11}(t)A_{22}(t) - A_{12}(t)A_{12}(t) \, ,
\label{eqd11}
\end{equation}
%
\begin{equation}
\Delta _{34}(t) = A_{33}^{[2]}(t) \odot A_{44}^{[2]}(t) -
A_{34}^{[2]}(t) \odot A_{34}^{[2]}(t) \, . \label{eqd12}
\end{equation}
In these expressions, $[\nabla_{\mathbf{q}}
\omega_{\mathbf{q}}\nabla_{\mathbf{q}} \omega_{\mathbf{q}}]$ denotes
the second order tensor with components $\partial \omega /\partial
q_{i}\partial \omega /\partial q_{j}$, while $F^{[2]} \odot G^{[2]}
= \sum \limits_{ij}F_{ij}G_{ji}$, with $\odot$ standing for full
contracted description.

\end{document}